\documentclass[lettersize,journal]{IEEEtran}
\usepackage{amsmath,amsfonts}
\usepackage{algorithmic}
\usepackage{algorithm}
\usepackage{array}
\usepackage[caption=false,font=normalsize,labelfont=sf,textfont=sf]{subfig}
\usepackage{textcomp}
\usepackage{stfloats}
\usepackage{url}
\usepackage{verbatim}
\usepackage{graphicx}
\usepackage{cite}
\begin{document}
\title{Scalable Air-to-Ground Wireless Channel Modeling Using Environmental Context and Generative Diffusion}

\author{Jingyi Tian,~\IEEEmembership{Graduate Student Member,~IEEE,} Lin Cai,~\IEEEmembership{Fellow,~IEEE,}

\thanks{Parts of this paper were submitted to the IEEE International Conference on Communications (ICC) 2026 and are currently under review \cite{my}.
}
\thanks{Jingyi Tian and Lin Cai are with the Department of Electrical and Computer Engineering, University of Victoria, Victoria V8P 5C2, Canada (e-mail: jingyit@uvic.ca, cai@uvic.ca).}
}



\maketitle

\begin{abstract}
The fast motion of Low Earth Orbit (LEO) satellites causes the propagation channel to vary rapidly, and its behavior is strongly shaped by the surrounding environment, especially at low elevation angles where signals are highly susceptible to terrain blockage and other environmental effects. Existing studies mostly rely on assumed statistical channel distributions and therefore ignore the influence of the actual geographic environment. In this paper, we propose an environment-aware channel modeling method for air-to-ground wireless links. We leverage real environmental data, including digital elevation models (DEMs) and land cover information, together with ray tracing (RT) to determine whether a link is line-of-sight (LOS) or non-line-of-sight (NLOS) and to identify possible reflection paths of the signal. The resulting obstruction and reflection profiles are then combined with models of diffraction loss, vegetation absorption, and atmospheric attenuation to quantitatively characterize channel behavior in realistic geographic environments. Since RT is computationally intensive, we use RT-generated samples and environmental features to train a scalable diffusion model that can efficiently predict channel performance for arbitrary satellite and ground terminal positions, thereby supporting real-time decision-making. In the experiments, we validate the proposed model with measurement data from both cellular and LEO satellite links, demonstrating its effectiveness in realistic environments.
\end{abstract}


\section{Introduction}
\IEEEPARstart{L}ow Earth Orbit (LEO) satellites operate at relatively low altitudes, which significantly reduces propagation delay. Multiple satellites can further cooperate in constellations to provide global coverage, making LEO systems a key enabler of next-generation ubiquitous connectivity. However, their low orbital height also leads to high relative velocities with respect to ground user terminals (UTs), so the satellite–user geometry evolves rapidly and the propagation channel becomes highly time-varying. As a result, link availability, capacity, and quality can change quickly, especially when the link is maintained at low elevation angles where the line-of-sight (LOS) between the satellite and the UT passes close to the horizon and is thus prone to severe blockage and frequent transitions to non-line-of-sight (NLOS) conditions.

Under such highly time-varying conditions, the surrounding environment becomes a key factor in determining communication performance. At low elevation angles, the propagation path is more likely to intersect terrain, buildings, and vegetation, which can obstruct, attenuate, or scatter the signal, especially in the Ku/Ka bands that are widely used in LEO satellite communications and are highly sensitive to propagation impairments \cite{itu-2108}. As a result, accurately capturing environmental effects on propagation is essential for the efficient and reliable operation of LEO satellite networks.

\begin{figure}
    \centering
    \includegraphics[width=1\linewidth]{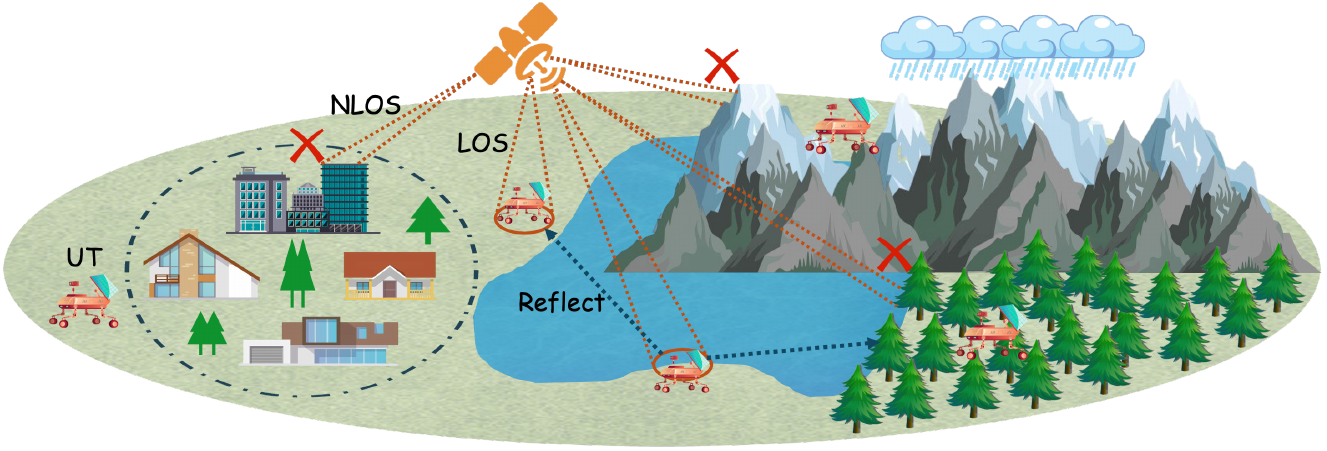}
    \caption{Environment Illustration}
    \label{el}
    \vspace{-4mm}
\end{figure}

Fig. \ref{el} illustrates how environmental factors such as mountains, vegetation, sea surfaces, and buildings shape channel behavior. For some UT–LEO links, signals may be obstructed by mountains, vegetation, or buildings, leading to NLOS conditions with severe degradation, especially at low elevation angles. Strong reflectors like sea surfaces induce multipath, while rainfall and clouds further impair link quality. Only when no obstruction exists between the satellite and the UT is the link considered LOS. The complexity of the environment highlights the need for an environment-aware channel modeling method capable of capturing diverse influences on air-to-ground links.

Ray tracing (RT) \cite{rt} is a deterministic channel modeling technique that computes electromagnetic propagation paths based on geometric optics, thereby reproducing signal propagation characteristics close to real conditions. It has been widely used in channel modeling. However, most existing RT-based methods still rely on distributional assumptions, such as treating obstruction as a random variable \cite{k,ricean} or tuning abstract parameters to emulate different environments \cite{ele-dis}, leaving the actual environmental information explicitly unmodeled. Measurement-based approaches \cite{nist} better reflect reality but are costly and impractical in inaccessible regions such as polar areas.

To systematically evaluate these environment-induced effects, we choose the Canadian Arctic as our region of interest (ROI), which features diverse terrain (–109.9 m to 5083.8 m) \cite{dem} with mountain ridges, dense forests causing obstruction and absorption, strong reflectors such as sea and permanent ice surfaces, as well as residential buildings and barren lands, thereby offering a comprehensive set of environmental factors for evaluation. Building on this ROI, we take two steps to address the gap that LEO satellite channel models rarely account for the spatially varying environment along the rapidly changing satellite–UT geometry.

First, we construct a RT-based model using digital elevation model (DEM) data \cite{dem}, in which each pixel represents terrain elevation, together with land cover data \cite{lc} to determine NLOS links at the pixel level and record obstacles. We compute terrain parameters (slope, aspect, roughness, curvature) and combine them with land cover to characterize surface reflectivity. We then quantify diffraction and extinction effects from terrain and vegetation obstructions, and multipath propagation from reflections. Meteorological data \cite{weather} are also used to quantify rain, cloud, and atmospheric attenuation, ensuring comprehensive environmental consideration.

Second, across large regions, the combinations of UT positions with satellites at different altitudes, elevations, and azimuths are effectively infinite, and the high computational cost of RT makes direct calculation infeasible for real-time use. To tackle this, we first cluster the area by terrain, land cover, and functional types, then perform representative ground-point sampling within each cluster. Satellite altitude, elevation, and azimuth are likewise sampled, and RT is executed only for these sampled UT–satellite pairs. Based on these samples, we train a diffusion model \cite{ddpm} to predict channels at arbitrary locations and geometries, thereby avoiding large-scale, high-complexity RT computations. In summary, the main contributions of this paper are as follows:
\begin{itemize}
    \item \textbf{Environment‐Aware Channel Modeling}: We integrate terrain, land cover, and climate to determine their effects on signal propagation, enabling accurate air-to–ground link estimation.
    \item \textbf{Diffusion‐Based Prediction}: Channel performance is predicted from environmental inputs using a diffusion model, replacing costly RT and enabling adaptive scheduling.
    \item \textbf{Validation with Real Measurements}: Validation with real measured LEO satellite and cellular data demonstrates the model’s accuracy and practical applicability.
\end{itemize}

The remainder of this paper is organized as follows. Section II reviews related work on wireless channel modeling. Section III presents the proposed site-specific channel modeling methodology, including region clustering and geometric channel estimation. Section IV details the proposed diffusion-based prediction framework. Section V discusses the experimental setup and evaluation results, and Section VI concludes the paper with future directions.

\section{Related work}
Wireless channel modeling has been studied across ground, aerial, and satellite scenarios, evolving from purely statistical descriptions toward environment-aware methods. For ground-to-ground links, models are relatively mature. Recent work begins to incorporate environmental information and AI, for example, \cite{cost} combined building maps with RT-derived LOS information and uses neural networks to predict urban path loss. However, the scenario is restricted to flat city layouts. Terrain undulation and rural/remote environments are not considered, limiting generality beyond idealized urban grids. NIST’s long-running measurement programs \cite{nist} collected extensive real-world data in typical urban and indoor environments, providing valuable references for next-generation standards, but the campaigns are costly and geographically limited.

Air-to-ground and air-to-sea modeling for UAV links is more recent and remains less mature. In maritime environments, \cite{u2s} proposed a UAV–ship (U2S) MIMO channel model that incorporates sea clutter and wave shadowing combined with UAV 6D motion, but relies on Rayleigh-based fading assumptions and lacks large-scale measurement support. For UAV-to-ground communications, \cite{hua} developed an ultra-wideband nonstationary model that combines geometry and RT to generate time-varying multipath components and jointly capture large and small-scale fading, yet its scenario diversity is limited and its dependence on full RT makes real-time use and large-area deployment challenging. In the agricultural IoUT setting, \cite{iout} augmented traditional path-loss models with empirical terrain, crop, and weather corrections, the environmental factors are introduced only as coarse correction terms and are not validated against independent measurements, making it difficult to assess robustness in more complex or unseen environments.

Satellite-to-ground channel modeling is even more constrained, with most studies focusing on low-elevation degradation and often reverting to abstract statistics. Flexible distribution-based models, such as the K-distribution for shadowing in LEO direct-to-satellite links \cite{k} and the $\eta$–$\mu$ shadowed Ricean model for land-mobile satellite channels \cite{ricean}, can fit a range of fading behaviors but remain environment-agnostic and largely unvalidated by real measurements in diverse terrains. Hybrid approaches attempt to incorporate environment features through RT. \cite{ele-dis} constructed simplified urban building and vegetation layouts and then fits elevation-dependent distributions for dense city and residential areas, but still treats the ground as a flat plane and cannot represent realistic terrain complexity. Recent learning-based works push further on prediction accuracy. \cite{fas-llm} introduced a large language model (LLM)–based predictor for OTFS-enabled satellite–FAS links using historical channel coefficients, and \cite{dp} proposed a joint channel prediction and multibeam precoding method for LEO satellite IoT, trained on 3GPP derived simulated channel state information. These methods improve sequence prediction under long delays, yet they rely purely on synthetic or history-only channel coefficients without explicitly modeling terrain, land cover, or weather, raising concerns about generalization to real, heterogeneous environments.

Overall, despite substantial progress from ground to UAV and satellite scenarios, most existing models still insufficiently capture real environments: terrain variations are often ignored or oversimplified, statistical abstractions dominate, measurement validation is rare or confined to limited regions, and heavy reliance on full RT makes large-scale, real-time use impractical. Field campaigns demonstrate the value of real measurements but are expensive and difficult to replicate in inaccessible areas, especially polar regions.

To fill this gap, our work builds an explicitly environment-aware air–to–ground channel model driven by real geospatial information. We integrate DEM, land cover data, and meteorological records to construct a realistic propagation environment that simultaneously accounts for terrain variation, surface type, and atmospheric conditions. Within this environment, RT is applied only to a set of sampled UT–satellite geometries, after which a diffusion model predicts channel states over the entire region, avoiding repeated large-scale RT in real-time operation. The approach is validated against measured data, showing that it captures environmental effects with high fidelity while supporting real-time applications and extending to previously inaccessible regions such as the Arctic.

\section{Environment-Aware Channel Modeling}
\begin{figure}
    \centering
    \includegraphics[width=1\linewidth]{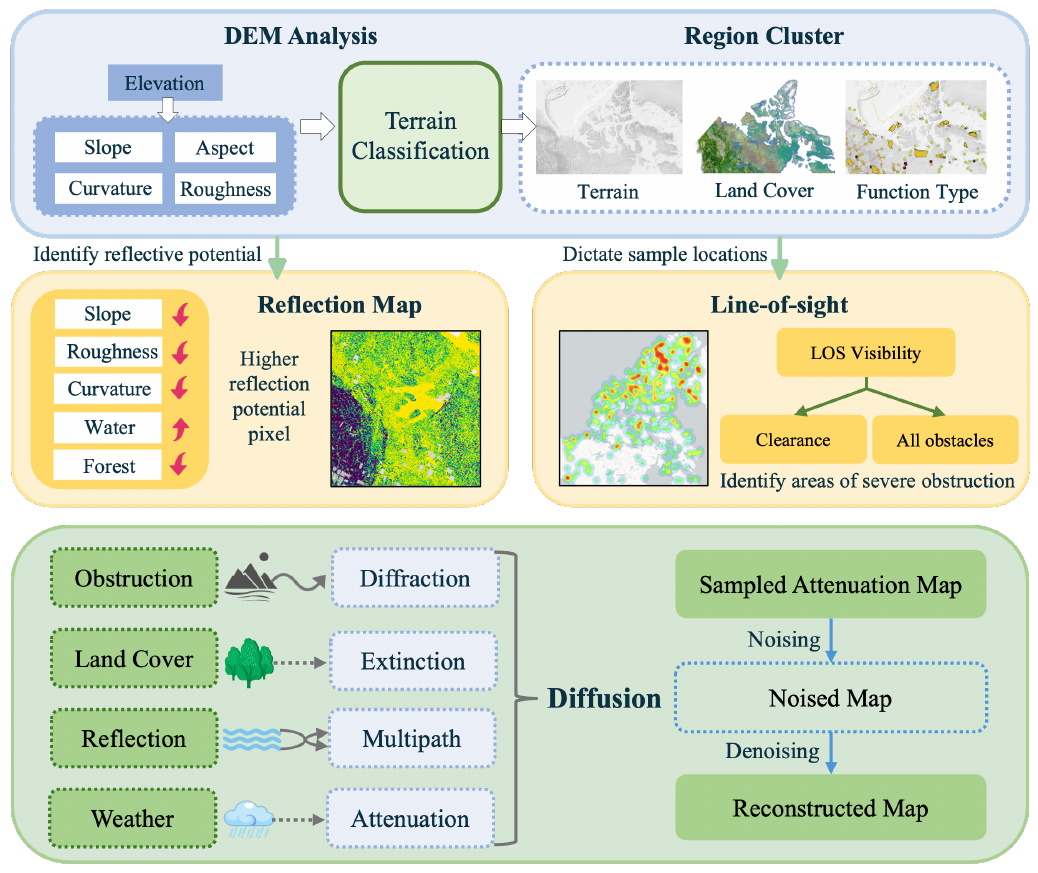}
    \caption{Methodology Framework}
    \label{arc}
\end{figure}
\subsection{Overview}
We propose an environment-aware air-to-ground channel modeling method and apply it in our ROI: Canada’s Arctic which comprises Yukon (YT), Northwest Territories (NT) and Nunavut (NU), aiming to capture the impacts of complex and diverse environments on LEO satellite channels. Fig. \ref{arc} shows the workflow of the proposed approach.

First, we extract terrain parameters such as slope from DEM data and apply the Weiss classification \cite{weiss} scheme to categorize the terrain. These terrain classes are then combined with land cover types and land-use functions (e.g., protected areas, mining zones, residential settlements) to perform clustering on the ROI. Given that direct computation of all satellite–UT links is infeasible, representative ground points—prioritizing high terrain elevations and ridge-dense areas—are sampled in conjunction with satellite states spanning a range of elevation angles, azimuths, and altitudes.

Next, terrain parameters and land cover to assess reflection capability of each pixel, producing a reflection map. With these information, we quantify signal loss for each link, including diffraction from terrain and vegetation obstructions, vegetation extinction, multipath from reflections, and atmospheric attenuation from meteorological data, forming a comprehensive channel attenuation model. 

Instead of performing computationally intensive RT simulations for every potential link at runtime, we train a diffusion-based predictor on sampled RT results and environmental features to estimate channel conditions at arbitrary locations, providing real-time inputs for scheduling decisions. The following sections detail each design.

\subsection{Region Cluster}
\subsubsection{Region Cluster}
The DEM data we used has a spatial resolution of 10 $\times$ 10 meters per pixel. Based on our calculations, the ROI contains approximately 109.5 billion pixels. Since satellites at different altitudes operate with varying elevation and azimuth angles, the number of possible link combinations between the satellites and UTs is effectively infinite, making exhaustive computation and storage impractical.
To ensure that the sampling points better reflect the characteristics of different environments instead of sampling randomly, we first cluster the entire region using environmental data:
\begin{itemize}
    \item Terrain features: We compute slope, aspect, roughness, curvature, Topographic Position Index (TPI), and Terrain Ruggedness Index (TRI) from the DEM using RichDEM \cite{rich-dem}, and apply the Weiss classification method \cite{weiss} to divide the terrain into valleys, ridges, plains, and other categories based on flatness and slope.
    \item Land cover features: The land cover data of ROI contains 19 categories, including urban built-up areas, forests, shrubs, grasslands, and water/permanent ice, etc., with an overall accuracy of 86.9\% \cite{lc}.
    \item Functional region features: We also use regional functional distribution data \cite{func-type}, which identifies the locations of residential areas, protected areas, mining regions, wilderness, etc., providing useful information about the relative density of UTs in different areas.
\end{itemize}

Based on these information, we apply k-means clustering \cite{kmeans} to the entire region to guide the sampling for LOS and reflection calculations, ensuring that the distribution of sampling points accounts for the diverse environmental conditions.

\subsubsection{Stratified sampling}
Let the cluster set be $\mathcal{C} = \{C_1, \ldots, C_m\}$, Cluster $C_i$ contains $N_i$ candidate points, and the total number of candidates is $N = \sum_{i=1}^{m} N_i$. To obtain samples that are both widely covered and representative under a fixed budget, we first allocate a sample quota to each cluster. The sampling weight of cluster $C_i$ is defined as:
\begin{equation}
W_i = \rho_i \cdot (\alpha \omega_T + \beta \omega_F + \gamma \omega_L + \delta \omega_E),
\end{equation}
where $\rho_i = N_i / N$ is the area (size) proportion of cluster $C_i$, $\omega_T$, $\omega_F$, and $\omega_L$ are the weighting factors for terrain type, functional region, and land cover, respectively, $\omega_E$ is the elevation gain factor, and the mixture coefficients satisfy $\alpha + \beta + \gamma + \delta = 1$. By tuning these coefficients, one can realize task-specific sampling strategies, like LOS analysis or reflection analysis. Given a total budget $S$, the quota assigned to cluster is $s_i = S \cdot \frac{W_i}{\sum_{j=1}^{m} W_j}$.

Within each cluster, we then select concrete samples according to its quota $s_i$. Define a feature combination $k = (t, f, l)$, where $t$, $f$, and $l$ represent the terrain class, functional region class, and land cover class, respectively. The weight of combination $k$ is:
\begin{equation}
w_k = \omega_T(t) \times \omega_F(f) \times \omega_L(l),
\end{equation}
and its allocated sample count is:
\begin{equation}
s_{i,k} = \max\left(s_{min}, \left\lfloor \frac{w_k}{\sum_{k'} w_{k'}} \cdot s_i \right\rfloor\right),
\end{equation}
where $k'$ denotes all feature combinations that exist within cluster $C_i$, $s_{min}$ enforces a minimum per-combination quota so that every combination receives samples.

Within each combination $k$, ground samples are selected uniformly at random, as points within the same combination share similar environmental characteristics. To prevent excessive spatial clustering, we impose a minimum distance constraint $d_{min}$ so that the distance between any two retained samples is at least $d_{min}$. This multi-stage stratified sampling strategy yields a sample set that (i) covers the main environment combinations, (ii) prioritizes more influential instances, and (iii) sufficiently spreads out in space.

In practice, the coefficients should be adjusted to the analysis goal. For LOS blockage analysis, the allocation favors the terrain class $\alpha$ and terrain elevation $\delta$, and and assigning larger $\omega_T$ to ridges/valleys while down-weighting water in $\omega_L$ to better capture ridges, valleys, and other relief features that create obstructions. For reflection-path analysis, the allocation favors the land cover $\gamma$ and boosting $\omega_L$ for highly reflective classes, emphasizing strongly reflective areas. This parameterized design allows the same framework to adapt to multiple satellite-communication analysis scenarios.

\subsection{LOS and Reflection Map}
In our ROI, the elevation values of DEM pixels range from $-109.9\,\text{m}$ to $5083.8\,\text{m}$, indicating substantial relief. At low and moderate elevation in a high relief region, satellite links are more susceptible to terrain obstruction. For Ku/Ka bands commonly used in satellite communications, vegetation can also introduce notable attenuation. Therefore, for each DEM pixel we augment the terrain elevation with a representative height of its overlying land cover to form the pixel’s \emph{effective terrain height}. At each ground sampling location, we further generate multiple satellite positions: elevation angles from $25^\circ$ to $85^\circ$ in $15^\circ$ steps; azimuth angles from $0^\circ$ to $300^\circ$ in $60^\circ$ steps; and three orbital altitudes of $500$\,km, $850$\,km, and $1200$\,km.

\begin{figure}
    \centering
    \includegraphics[width=0.9\linewidth]{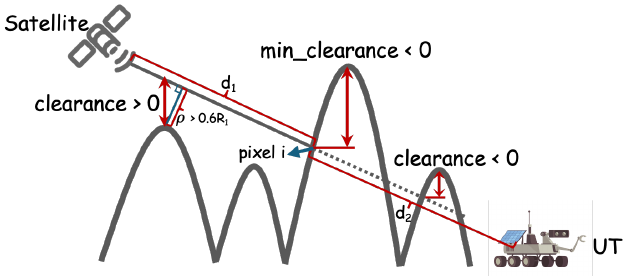}
    \caption{Path Profile}
    \label{pf}
\end{figure}

We implement a grid-based RT algorithm incorporating Earth-curvature correction and a Fresnel-zone criterion. To ensure computational efficiency over the large-scale ROI, a hierarchical traversal strategy is adopted. Since DEM products are typically provided as tiled datasets, the Liang-Barsky line-clipping algorithm \cite{lb} is first applied to rapidly identify the specific DEM data blocks intersected by the link vector. Subsequently, within each identified block, Bresenham’s line algorithm \cite{bresenham} is employed to discretize the propagation path, strictly determining the raster pixels traversing the link. At each sampled pixel $i$, we first correct the link height to account for Earth curvature, which together make the Earth appear flatter. In the 4/3-Earth model the effective radius is $R_{\mathrm{eff}} = 4/3 \times R_{\oplus}$ and $R_{\oplus}$ is the physical Earth radius. The corrected link height is:
\begin{equation}
    H_{\mathrm{link}} = H_{\mathrm{link}}^{raw} - \frac{d_1 d_2}{2\,R_{\mathrm{eff}}},
\end{equation}
where $d_1$, $d_2$ are the 3D distances from pixel $i$ on the link to the satellite and the UT, respectively. $H_{\mathrm{link}}^{raw}$ is the link height obtained by straight-line geometry between the UT and the satellite.

We compute the $clearance = H_{link} - H_{terrain}$ which corresponds to the margin between the link height and the effective terrain terrain height. The first-Fresnel-zone radius $R_1$, which represents the region around the link where most of the signal energy propagates. For adequate clearance, the obstacle’s normal distance to the link should be at least $0.6R_1$.
\begin{equation}
    R_1 = \sqrt{ \lambda \,\frac{d_1 d_2}{d_1 + d_2} },
\end{equation}
where $\lambda$ is the signal wavelength. Fig. \ref{pf} provides an intuitive illustration. The Fresnel clarity ratio is
\begin{equation}
    \rho = \frac{\rm clearance \times \cos(elevation)}{R_1},
\end{equation}
a negative value of the $\rm min\_clearance$ indicates that the link is obstructed. If $\rm min\_clearance \ge 0$ and $\rho \ge 0.6$, we classify the link as LOS; otherwise, it is NLOS.

\begin{algorithm}[t]
\caption{LOS/NLOS \& Path Profile Construction}
\label{alg:los-path-profile}
\begin{algorithmic}
\REQUIRE Coordinates of Start $P_s$ and End $P_e$ of a link; DEM Blocks $\mathcal{D}$; Wavelength $\lambda$; Effective Earth Radius $R_{\text{eff}}$

\ENSURE Output LOS/NLOS; path profile $\mathcal{P}$

\STATE $\mathcal{P} \gets \emptyset$; $\text{min\_clearance} \gets +\infty$; $\text{min\_}\rho \gets +\infty$
\STATE $\mathcal{B}_{\text{intersect}}\leftarrow \emptyset$; Pixel\_List $\leftarrow \emptyset$

\STATE /* Find intersected DEM blocks along the link */
\STATE $\mathcal{B}_{\text{intersect}} \leftarrow \text{LiangBarskyClipping}(P_s, P_e, \mathcal{D})$

\STATE /* Find pixels along link */
\FORALL{Block $b \in \mathcal{B}_{\text{intersect}}$}
    \STATE $L_{\text{local}} \leftarrow \text{BresenhamLine}(P_s, P_e, b)$
    \STATE Append $L_{\text{local}}$ to Pixel\_List
\ENDFOR

\FOR{pixel $i$ in Pixel\_List}
  \STATE Extract distance $d_1(i), d_2(i)$ and effective terrain height $H_{\text{terrain}}(i)$
  \STATE /* Earth-curvature corrected link height */
  \STATE $H_{\mathrm{link}}(i) \gets H_{\mathrm{link}}^{\mathrm{raw}}(i) - \dfrac{d_1(i)\,d_2(i)}{2R_{\mathrm{eff}}}$
  \STATE /* Clearance and Fresnel clarity */
  \STATE $\text{clearance} \gets H_{\mathrm{link}}(i) - H_{\mathrm{terrain}}(i)$
  \STATE $R_1 \gets \sqrt{\dfrac{\lambda\, d_1(i)\, d_2(i)}{d_1(i)+d_2(i)}}$
  \STATE $\rho \gets \dfrac{\text{clearance}\cdot \cos\theta}{R_1}$

  \STATE $\text{min\_clearance}\leftarrow{}\allowbreak
        \min(\text{min\_clearance},\text{clearance})$
  \STATE $\text{min\_}\rho\leftarrow\min(\text{min\_}\rho,\rho)$

  \IF{$\text{clearance} < 0$ \OR $\rho < 0.6$}
    \STATE append $(\text{dist}_h(i),\, H_{\mathrm{terr}}(i),\, \text{clearance})$ to $\mathcal{P}$
  \ENDIF
\ENDFOR

\IF{$\text{min\_clearance} \ge 0$ \AND $\text{min\_}\rho \ge 0.6$}
  \STATE \textbf{return} LOS, $\mathcal{P} = \emptyset$
\ELSE
  \STATE sort $\mathcal{P}$ by $\text{dist}_h$ ascending
  \STATE \textbf{return} NLOS, $\mathcal{P}$
\ENDIF

\end{algorithmic}
\end{algorithm}

As in Algorithm~\ref{alg:los-path-profile}, after classifying the link as LOS or NLOS, we construct a \emph{path profile} $\mathcal{P}$ for NLOS links: the obstructed pixels on NLOS paths are ordered by horizontal distance to form a profile curve, and we record the clearance together with the location and height of each obstacle point. This provides both a global LOS/NLOS decision and the locations and heights of obstacles along the path, which will be used subsequently for channel estimation.

In regions with abundant water and permanent ice, signal reflections are also non-negligible. Therefore, we first construct a reflection-capability raster based on the DEM and land cover data. As mentioned earlier, we compute slope $S$, roughness $R$, and curvature $K$ from the DEM. After normalization, the terrain-based reflection score $T$ is defined as:
\begin{equation}
\end{equation}
where $w_S$, $w_R$, and $w_K$ are the respective weights of slope, roughness and curvature. In this way, flat (low-slope) and smooth (low-roughness) areas receive higher scores, and concave regions (negative curvature) are further enhanced.  

Next, land cover information is incorporated by assigning a linear attenuation coefficient $\beta_\ell$ to each type. The final reflection coefficient is given by:
\begin{equation}
R = T \cdot (1 + \beta_\ell).
\end{equation}

In our settings, water surfaces are assigned $\beta_{\text{water}}=+0.50$ to enhance reflection, while forests are assigned $\beta_{\text{forest}}=-0.10$ to attenuate reflection. The resulting reflection map provides a quantitative measure of spatial reflection potential, where larger values indicate stronger reflective capability, serving as input for multipath modeling.

\subsection{Channel Estimation}
In this section, we quantify channel losses from environmental effects: climatic, obstruction, and multipath. For climatic effects, we apply ITU-R P.618 \cite{itu-618} using meteorological data \cite{weather} to model rain, cloud, and atmospheric attenuation, and incorporate these into the channel model. For obstruction effects, we consider two components: diffraction loss from terrain and land cover, and absorption loss from vegetation. Using the NLOS obstacle path profile, we apply the Delta-Bullington model \cite{itu-452} to compute diffraction loss. The model first determines obstacle positions and clearance heights, then sequentially evaluates diffraction loss for each obstacle with ITU formulas, and finally transform multiple obstacles into an equivalent single obstacle profile, on which the total diffraction loss of the link is calculated.

For vegetation loss, we sample along the path to reduce computational cost. Computing vegetation loss first requires the penetration distance through vegetation and the vegetation type. The challenge is to correctly identify the entry and exit of each vegetation segment. At each sampled point along the path we test the land cover type. To refine each entry/exit more accurately, we apply a bisection search, as described in Algorithm~\ref{alg:bisect-boundary}: if the current sample and its left neighbor share the same land-cover type, the boundary is on the left; if it matches the right neighbor, it is on the right. For vegetation segments $\mathcal{V} = \{[s_{\mathrm{in}}^{(k)}, s_{\mathrm{out}}^{(k)}, \text{class}^{(k)}]\}_k$ recorded along the path, where $k=1,2,\ldots$ indexes segments. We compute the \emph{penetration slant distance} as $s_{\mathrm{out}}-s_{\mathrm{in}}$, where $s$ denotes the cumulative \emph{slant} path distance along the link, which is then passed to ITU-R P.833 model~\cite{itu-833} with parameters adapted to frequency and land cover type.
\begin{algorithm}[t]
\caption{Vegetation Segmentation Detection}
\label{alg:bisect-boundary}
\begin{algorithmic}
\REQUIRE Input path distance $s\in[0,S_{\max}]$; land-cover oracle $\mathrm{Type}(s)$; sampling step $\Delta s$; bisection tolerance $\varepsilon$
\ENSURE Output vegetation segments $\mathcal{V} = \{[s_{\mathrm{in}}^{(k)}, s_{\mathrm{out}}^{(k)}, \text{class}^{(k)}]\}_k$

\STATE $\mathcal{V}\gets\emptyset$; $s\gets 0$; $T_{\mathrm{prev}}\gets \mathrm{Type}(0)$; $\text{open}\gets \mathrm{false}$

\WHILE{$s < S_{\max}$}
  \STATE $s_{\mathrm{next}} \gets \min(s+\Delta s,\ S_{\max})$; $T \gets \mathrm{Type}(s_{\mathrm{next}})$
  \IF{$T \neq T_{\mathrm{prev}}$}
    \STATE /* Boundary in $(s,\,s_{\mathrm{next}})$ by bisection */
    \STATE $L\gets s$; $R\gets s_{\mathrm{next}}$
    \WHILE{$R-L > \varepsilon$}
      \STATE $M\gets (L+R)/2$; $T_M\gets \mathrm{Type}(M)$
      \IF{$T_M = \mathrm{Type}(L)$}      
        \STATE $R\gets M$
      \ELSIF{$T_M = \mathrm{Type}(R)$}  
        \STATE $L\gets M$
      \ELSE                             
        \STATE $R\gets M$
      \ENDIF
    \ENDWHILE
    \STATE $s^\star \gets (L+R)/2$; \\
    $T^\star \gets \mathrm{Type}(s^\star)$

    \IF{$T^\star$ is vegetation \AND $\neg \text{open}$}
      \STATE /* ENTRY */
      \STATE $s_{\mathrm{in}}\gets s^\star$; $\text{class}\gets T^\star$; $\text{open}\gets \mathrm{true}$
    \ELSIF{$T^\star$ is non-vegetation \AND \text{open}}
      \STATE /* EXIT */
      \STATE $s_{\mathrm{out}}\gets s^\star$
      \STATE append $[s_{\mathrm{in}},\, s_{\mathrm{out}},\, \text{class}]$ to $\mathcal{V}$
      \STATE $\text{open}\gets \mathrm{false}$
    \ENDIF
  \ENDIF
  \STATE $T_{\mathrm{prev}}\gets T$; $s \gets s_{\mathrm{next}}$
\ENDWHILE

\IF{\text{open}}
  \STATE /* Path ends inside vegetation */
  \STATE append $[s_{\mathrm{in}},\, S_{\max},\, \text{class}]$ to $\mathcal{V}$
\ENDIF
\RETURN $\mathcal{V}$
\end{algorithmic}
\end{algorithm}

For multipath effects, enumerating all reflections is prohibitive. We introduce a trade-off by sampling a circle around the UT with distance-dependent density, as LEO satellites transmit signals in beams. Reflection potential at sampled points is evaluated from the reflection map, and only high-potential candidates are retained. Among these, paths are validated by requiring the incident angle to equal the reflection angle. The TWDP model \cite{twdp} is then applied to these validated paths, from which the direct path and the strongest reflection are selected to derive overall multipath gain/fading. Combined with free-space loss and other quantified effects, these results form the overall channel behavior.

\section{Diffusion-Based Prediction}
Exhaustively modeling the near-infinite geometric combinations of ground points and moving satellites is computationally infeasible, while high-fidelity RT simulations remain too costly for real-time applications. To address this, we employ diffusion models, which fundamentally operate by learning a reverse denoising process—corrupting data structures with gradual noise during training and then learning to iteratively recover complete samples from noisy or incomplete states. This generative mechanism of restoring order from disorder is inherently suited for reconstructing full fields from partial observations, aligning perfectly with our scenario. Specifically, we treat the environment map as a multi-channel image where limited channel loss observations are retained only at selected pixels, and we leverage the diffusion model’s capability in modeling spatial correlations to reconstruct the complete channel information across the entire region, conditioned on these sparse inputs.

\subsection{Input Representation}
We represent a region as a 3D tensor $X \in \mathbb{R}^{C \times H \times W}$, where $C$ is the number of channels and $H, W$ are spatial dimensions. The channels combine terrain, land-cover, and geometry:
\[
C = \big[\text{DEM},\text{Slope},\text{Aspect},\text{LC\_emb}\big],
\]

The first four channels come from DEM, slope, aspect, and land cover index. Since land over index is an integer label which carries no semantic information, using it directly may induce spurious correlations. Instead, for each pixel with class id $c \in \{0,\ldots,K-1\}$, we map $c$ to an $E$-dimensional continuous vector via a learnable embedding matrix
$W \in \mathbb{R}^{K \times E}$. The $E$ components of embedded vector are then expanded along the channel dimension, forming $\text{LC\_emb}_{0..E-1} \;\in\; \mathbb{R}^{E \times H \times W}$,
which are optimized end-to-end with the training loss. This allows the model to learn semantic proximity between classes under varying terrain and geometric conditions (e.g., grassland closer to shrubland), improving cross-region generalization. The DEM undergoes robust Min-Max scaling based on the $1st$ and $99th$ percentiles to suppress outliers, slope is linearly mapped to the $[0, 1]$, while aspect is encoded via sine-cosine embeddings to resolve angular periodicity. Since the satellite geometric parameters are not pixel-level features, we combine them with the diffusion time-step embedding, and inject this joint embedding into every layer of the U-Net \cite{u-net} via residual blocks.

\subsection{Channel Loss Normalization}
We consider channel conditions across various elevation angles. At high elevations, the excess loss for the majority of LOS links approaches zero. Meanwhile, in high-frequency satellite communication, obstacles cause loss values to instantaneously span several orders of magnitude. Consequently, the observations exhibit a distinct mixed distribution characterized by \textit{zero-inflation} and \textit{heavy tails}. In this context, standard linear normalization would be skewed by a few extreme deep-fading values, causing the dominant near-zero features to be compressed into an indistinguishably narrow numerical range, thereby hindering the model's ability to capture subtle fluctuations in LOS scenarios. Conversely, logarithmic scaling is undefined at zero and fails to smoothly adapt to the sharp transitions from LOS to NLOS conditions.

To address this, we design a transformation scheme based on the Robust Inverse Hyperbolic Sine (Asinh) for the standardization of channel loss observations. The Asinh facilitates a smooth non-linear transition: in the low-loss regime ($m \to 0$), the function exhibits approximately linear behavior ($\operatorname{asinh}(m) \approx m$), thereby meticulously preserving fine-grained details near the zero point; whereas in the high-loss regime ($m \to \infty$), it demonstrates logarithmic characteristics ($\operatorname{asinh}(m) \approx \ln(2m)$), compressing the heavy-tailed distribution associated with deep fading. Consequently, this design significantly enhances both the training stability and convergence rate of the diffusion model in complex hybrid scenarios.

First, the raw channel loss observations $m_{\text{dB}}$ are converted into linear power ratios $m_{\text{lin}} = 10^{m_{\text{dB}}/10}$. Subsequently, to align the input distribution with Asinh's linear-to-logarithmic transition, thereby simultaneously preserving subtle features near zero and compressing extreme values associated with deep fading, we introduce a scaling factor $\eta$ to define the scaled linear power ratio $\tilde{m} = m_{\text{lin}} / \eta$, which is subsequently processed by the inverse hyperbolic sine transformation to obtain the intermediate variable $n$:
\begin{equation}
    n = \operatorname{arcsinh}\left(\tilde{m} \right),
\end{equation}
where $\eta$ is set to the $80^{\text{th}}$ percentile of the linear data distribution, serving to adaptively determine the transition point between the linear and logarithmic regions. Finally, to mitigate the impact of outliers on the statistical distribution, we employ $Z-Score$ standardization to obtain the final normalized channel observation $z$:
\begin{equation}
    z = \frac{n - \mu}{\sigma},
\end{equation}
where $\mu$ and $\sigma$ represent the robust mean (the median of $n$) and the robust standard deviation (derived from the Interquartile Range of $n$), respectively.

\subsection{Dual-Head U-Net Denoiser}
Given $X$ and partially observed channel-loss field $\mathbf{x}_0 \in \mathbb{R}^{1\times H\times W}$, we synthesize a noisy sample at timestep $t$:

\begin{equation}
\;\mathbf{x}_t \;=\; \sqrt{\bar{\alpha}_t}\,\mathbf{x}_0 \;+\; \sqrt{1-\bar{\alpha}_t}\,\boldsymbol{\epsilon}, 
\boldsymbol{\epsilon}\sim\mathcal{N}(\mathbf{0},\mathbf{I}), t=1\!\to\!T,
\label{eq:forward}
\end{equation}
$\bar{\alpha}_t$ is a predefined coefficient from the noise schedule that sets the signal–noise mixture at step $t$. $\boldsymbol{\epsilon}$ is standard Gaussian white noise with the same shape as \(\mathbf{x}_0\).

\begin{figure*}
    \centering
    \includegraphics[width=1.0\linewidth]{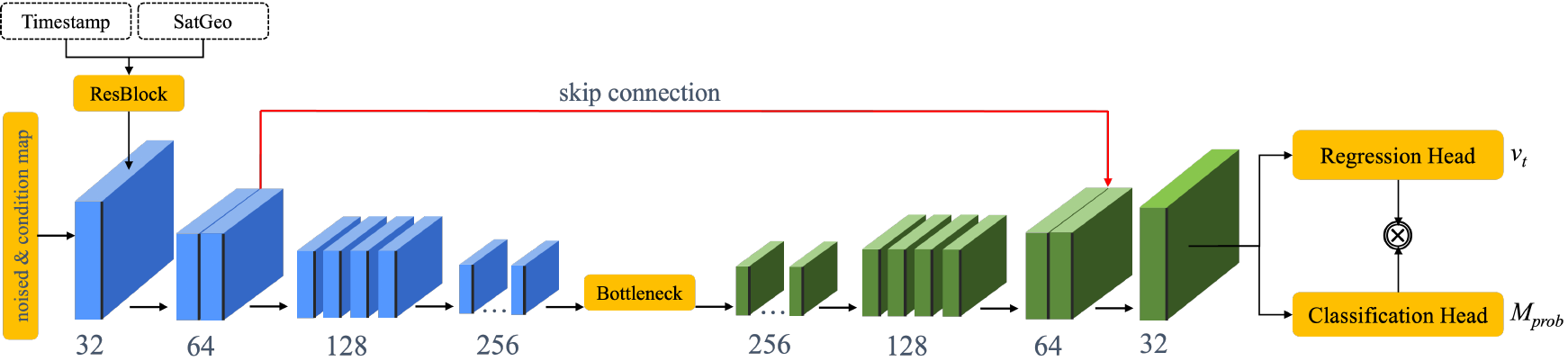}
    \caption{Dual-Head U-Net}
    \label{unet}
\end{figure*}

We feed noised target, conditioning map and diffusion timestamp $(\mathbf{x}_t, t, X)$ into a U-Net denoiser $f_{\theta}$. The U-Net adopts an encoder–bottleneck–decoder architecture. Along the encoder path, residual blocks and strided convolutions repeatedly downsample the feature maps from 32 channels at the original resolution to 64, 128, and 256 channels at progressively lower spatial resolutions. This downsampling enlarges the receptive field and allows the network to capture broad spatial context at a manageable computational cost. The decoder then mirrors this process by upsampling the low-resolution features back to the original resolution (256 $\longrightarrow$ 128 $\longrightarrow$ 64 $\longrightarrow$ 32 channels), which enables precise, per-pixel predictions. Skip connections link encoder and decoder blocks at the same scale, so that high-resolution features from the contracting path are reused in the expanding path. 

When the channel is in LOS propagation state, its theoretical excess loss value should strictly be $0\text{ dB}$. However, due to the aforementioned \textit{zero-inflation} issue, single-head regression models often struggle to converge to a perfect zero during the final stage of denoising. This tendency results in residual non-zero values in unobstructed regions, creating background noise artifacts and blurring the boundaries of obstacles. To further address this challenge, we propose a \textit{Dual-Head U-Net} architecture as shown in Fig. \ref{unet}. Both task heads share the same encoder-decoder backbone to extract generalized environmental features but bifurcate into two independent branches at the final output layer:
\begin{enumerate}
    \item \textbf{Regression Head:} Outputs the continuous denoising target, responsible for quantifying the magnitude of the channel loss.
    \item \textbf{Classification Head:} Outputs pixel-wise binary logits. Following a Sigmoid activation, this yields a blockage probability map $M_{\text{prob}} \in [0, 1]$, which is used to explicitly distinguish between LOS and NLOS states.
\end{enumerate}

These two distinct task heads are coupled at the pixel level through a \textit{soft-gating mechanism}. The final channel loss field $X_{\text{final}}$ is formulated as:

\begin{equation}
    X_{\text{final}} = X_{\text{raw}} \odot (M_{\text{prob}})^\gamma,
\end{equation}
where $X_{\text{raw}}$ denotes the raw output of the regression head, $\odot$ represents the Hadamard product, and $\gamma$ is the gating intensity coefficient. Through this mechanism, the classification head functions as a learnable filter: in LOS regions where $M_{\text{prob}} \to 0$, the regression output is suppressed to zero; conversely, in obstructed regions where $M_{\text{prob}} \to 1$, the regression value is preserved. This strategy not only eliminates background artifacts but also significantly sharpens the boundaries of obstacles. Conditioned on both $t$ and the satellite geometry (SatGeo) at every resolution level, the network is tasked with denoising. To enhance the numerical stability of geometric data reconstruction, we transition from the standard noise prediction target ($\epsilon$-prediction) to \textit{v-parameterization}. In this formulation, we define the ground truth velocity target $v_t$ in the latent space as a linear combination of the clean signal $\mathbf{x}_0$ and noise $\epsilon$:

\begin{equation}
    v_t \equiv \alpha_t \epsilon - \sigma_t \mathbf{x}_0,
\end{equation}

Unlike $\epsilon$-prediction, v-parameterization provides consistent gradient estimates across the entire diffusion trajectory. The network $f_\theta$ is trained to approximate $v_t$. The training objective minimizes the loss function:
\begin{equation}
    \mathcal{L} = \mathbb{E}_{t, \mathbf{x}_0, \epsilon} \left[ \lambda_t \| v_t - f_\theta(\mathbf{x}_t, t, X) \|^2 + \lambda_{cls} \mathcal{L}_{cls} \right],
\end{equation}
where $\mathcal{L}_{cls}$ denotes the binary cross-entropy loss of the classification head. The total objective comprises two terms: a regression loss weighted by the time-dependent coefficient $\lambda_t$, and the classification loss scaled by the hyperparameter $\lambda_{cls}$.

\subsection{Inference}
In the inference phase, the model initiates from pure Gaussian noise $\mathbf{x}_T \sim \mathcal{N}(\mathbf{0}, \mathbf{I})$ and iteratively recovers the signal over $t = T \to 1$ using a DDIM sampler \cite{ddim}. We enforce a \textit{Physics-Informed Inpainting Strategy} to ensure the reconstructed field remains strictly anchored to the sparse site-specific ground truth.

At each timestep $t$, the network predicts the velocity field $f_\theta(\mathbf{x}_t, t, X)$. Under the v-parameterization scheme, we first estimate the clean signal $\hat{\mathbf{x}}_0$ from the current noisy state $\mathbf{x}_t$ via the linear transformation:
\begin{equation}
    \hat{\mathbf{x}}_0(\mathbf{x}_t, t, X) = \sqrt{\bar{\alpha}_t} \mathbf{x}_t - \sqrt{1 - \bar{\alpha}_t} f_\theta(\mathbf{x}_t, t, X).
    \label{eq:x0_recon}
\end{equation}

Based on this $\hat{\mathbf{x}}_0$, the DDIM sampler proposes a denoised latent state $\mathbf{x}_{t-1}^{\text{pred}}$ for the next timestep. However, directly retaining this prediction allows deviations from the observed physical reality. Since the intermediate state $\mathbf{x}_{t-1}$ resides in a high-noise manifold, inserting noise-free data creates a severe SNR discontinuity at the boundaries. This distribution mismatch destabilizes the U-Net, leading to incoherent artifacts. To resolve this, we perform a re-noising and masking operation:

\begin{enumerate}
    \item \textbf{Re-noising:} We project the sparse clean observations $\mathbf{x}_0^{\text{obs}}$ into the same noise manifold as the current timestep $t-1$ by adding Gaussian noise matching the schedule intensity, ensuring statistical consistency with the surrounding generated regions:
    \begin{equation}
        \mathbf{x}_{t-1}^{\text{known}} = \sqrt{\bar{\alpha}_{t-1}} \mathbf{x}_0^{\text{obs}} + \sqrt{1 - \bar{\alpha}_{t-1}} \mathbf{\epsilon}, \quad \mathbf{\epsilon} \sim \mathcal{N}(\mathbf{0}, \mathbf{I})
    \end{equation}
    
    \item \textbf{Masking:} Using the binary observation mask $M$, we fuse the model's prediction with the re-noised ground truth to obtain the final state for the next iteration:
    \begin{equation}
        \mathbf{x}_{t-1} = \mathbf{x}_{t-1}^{\text{pred}} \odot (1 - M) + \mathbf{x}_{t-1}^{\text{known}} \odot M.
    \end{equation}
\end{enumerate}

Through this iterative correction, accurate physical information effectively diffuses into unobserved regions without introducing boundary artifacts, guiding the model to synthesize a full-field channel loss map that is both geometrically consistent and physically accurate.

\section{Evaluation}
\subsection{Evaluation on Real Datasets}
\begin{figure}[htbp]
    \centering
    \subfloat[Obstruction rate.]{%
        \includegraphics[width=0.23\textwidth]{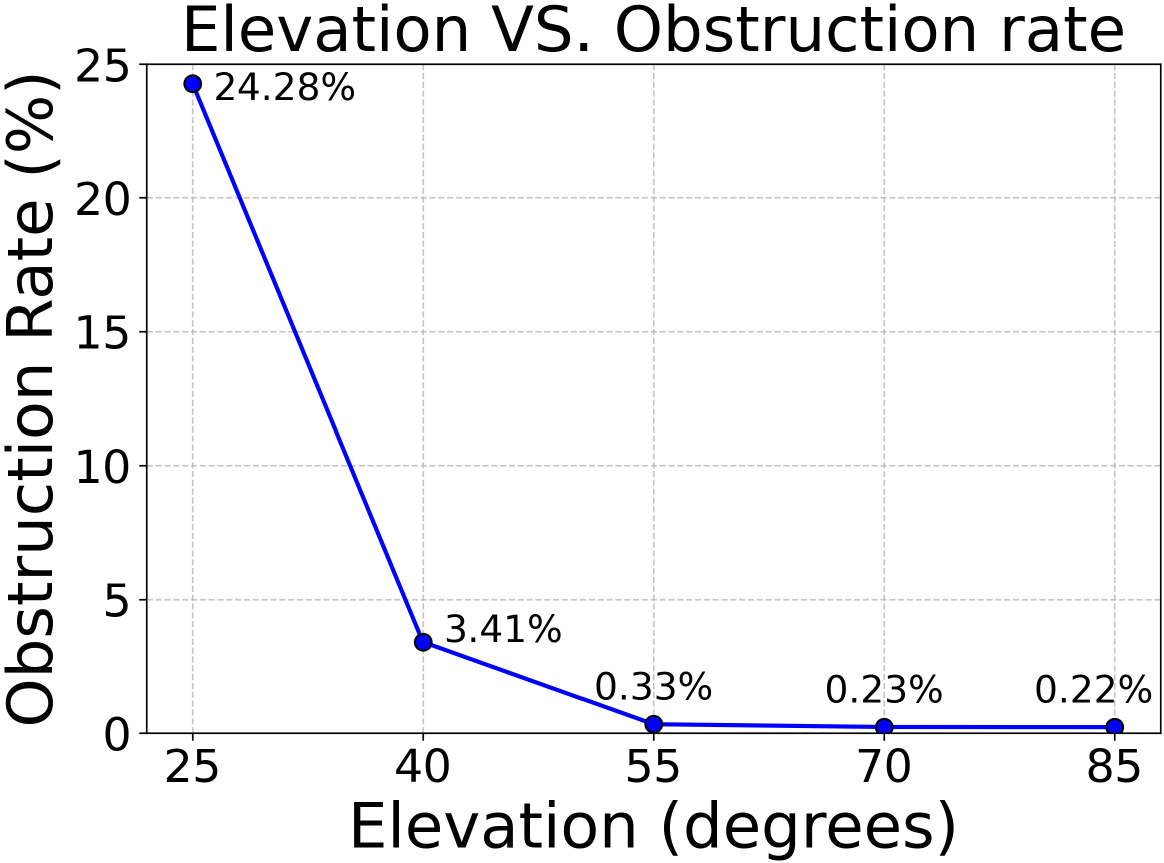}%
        \label{fig:obs_rate}%
    }
    \hfil
    \subfloat[Reflection map.]{%
        \raisebox{1.5ex}{%
            \includegraphics[width=0.22\textwidth]{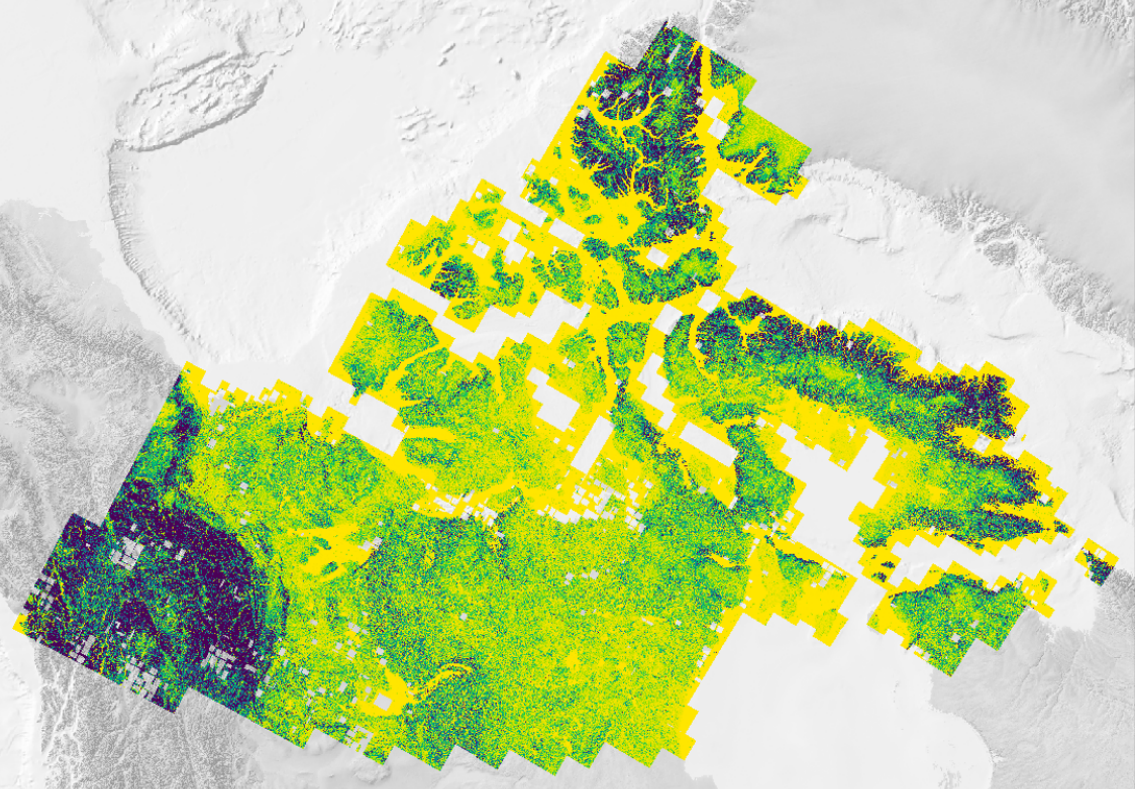}%
        }%
        \label{fig:reflection_map}%
    }
    \caption{Obstruction rate and reflection map.}
    \label{fig:pbs_rm}
\end{figure}

Using real datasets \cite{dem, lc, weather} and our method, we computed obstruction, reflection, and attenuation maps for the ROI. Fig. \ref{fig:obs_rate} shows obstruction rates at different elevations: they stay low at high elevations, both due to the natural advantage of elevation and the diluting effect of extensive flat plains, while rates are higher in the northeast with rugged ice terrain and in the southwest with dense forests and residential areas. The rate rises sharply at lower elevations, reaching 24.28\% at 25° even with the presence of extensive plains, which significantly impacts link reliability. Fig. \ref{fig:reflection_map} shows the reflection map, with weaker reflection in southwest forests and northeast rugged terrain but stronger in the central flat area, consistent with the obstruction distribution.

\begin{figure}[htbp]
    \centering
    \subfloat[$25^\circ$]{%
        \includegraphics[width=0.15\textwidth]{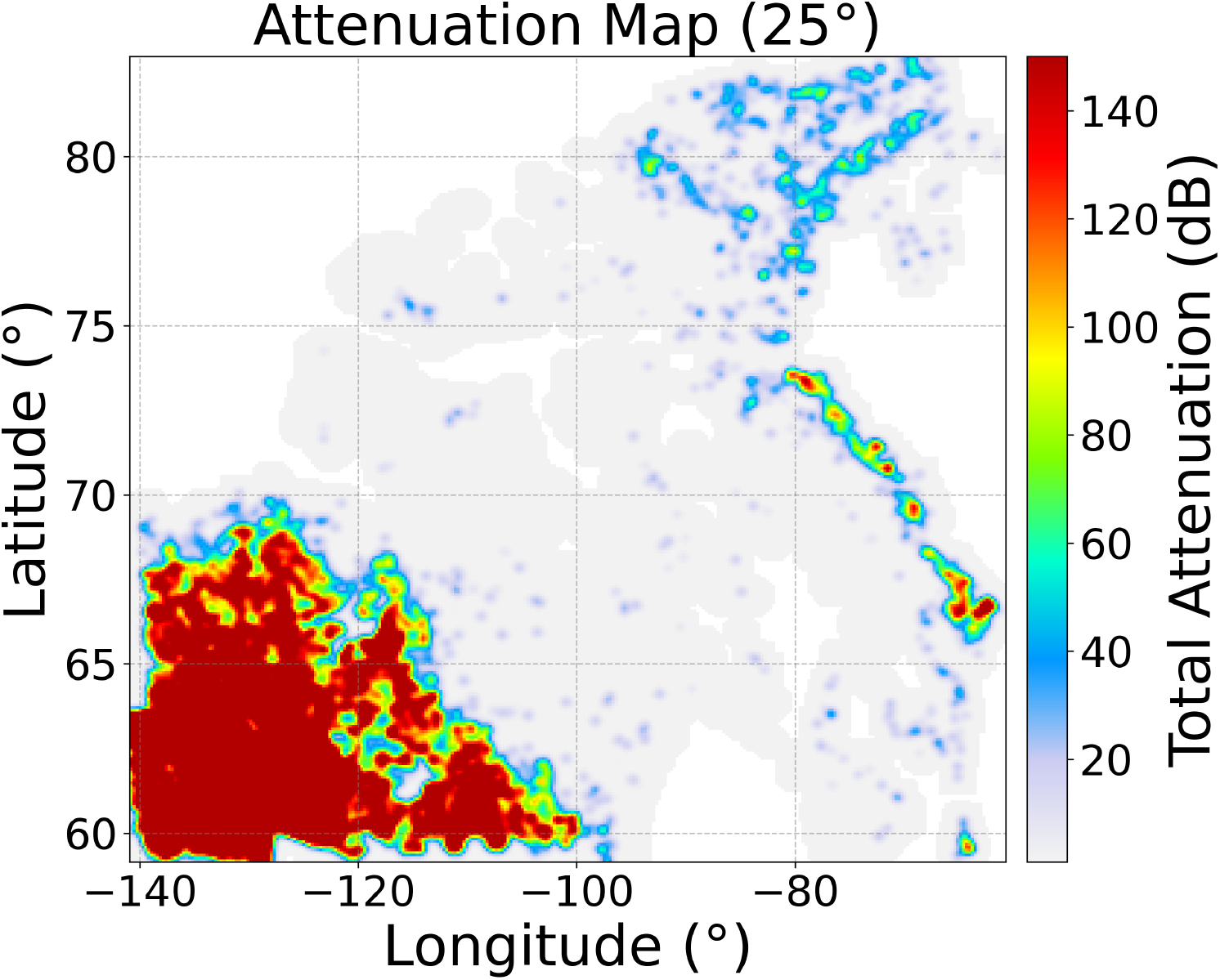}%
        \label{loss1}%
    }\hfil
    \subfloat[$40^\circ$]{%
        \includegraphics[width=0.15\textwidth]{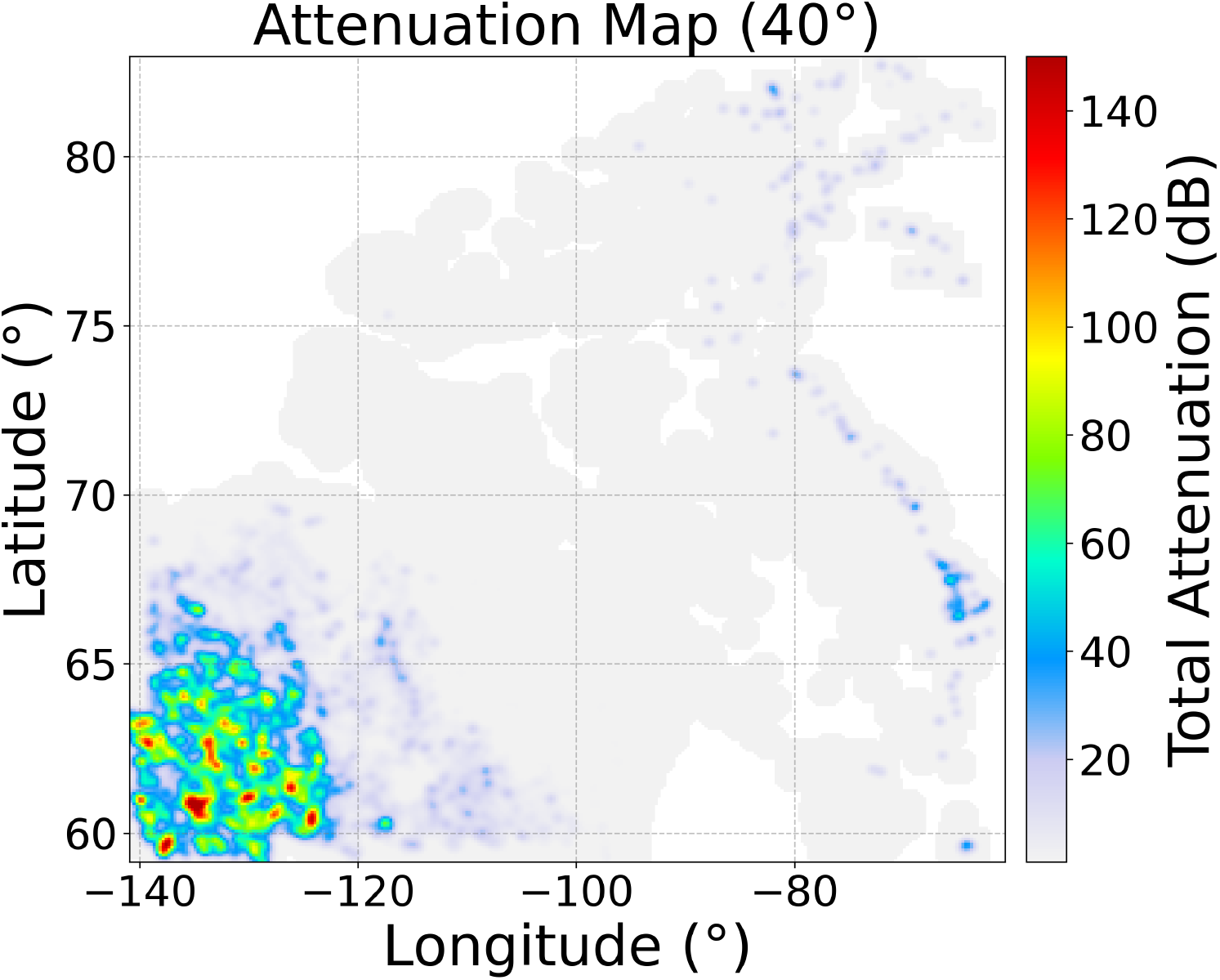}%
        \label{loss2}%
    }\hfil
    \subfloat[$55^\circ$]{%
        \includegraphics[width=0.15\textwidth]{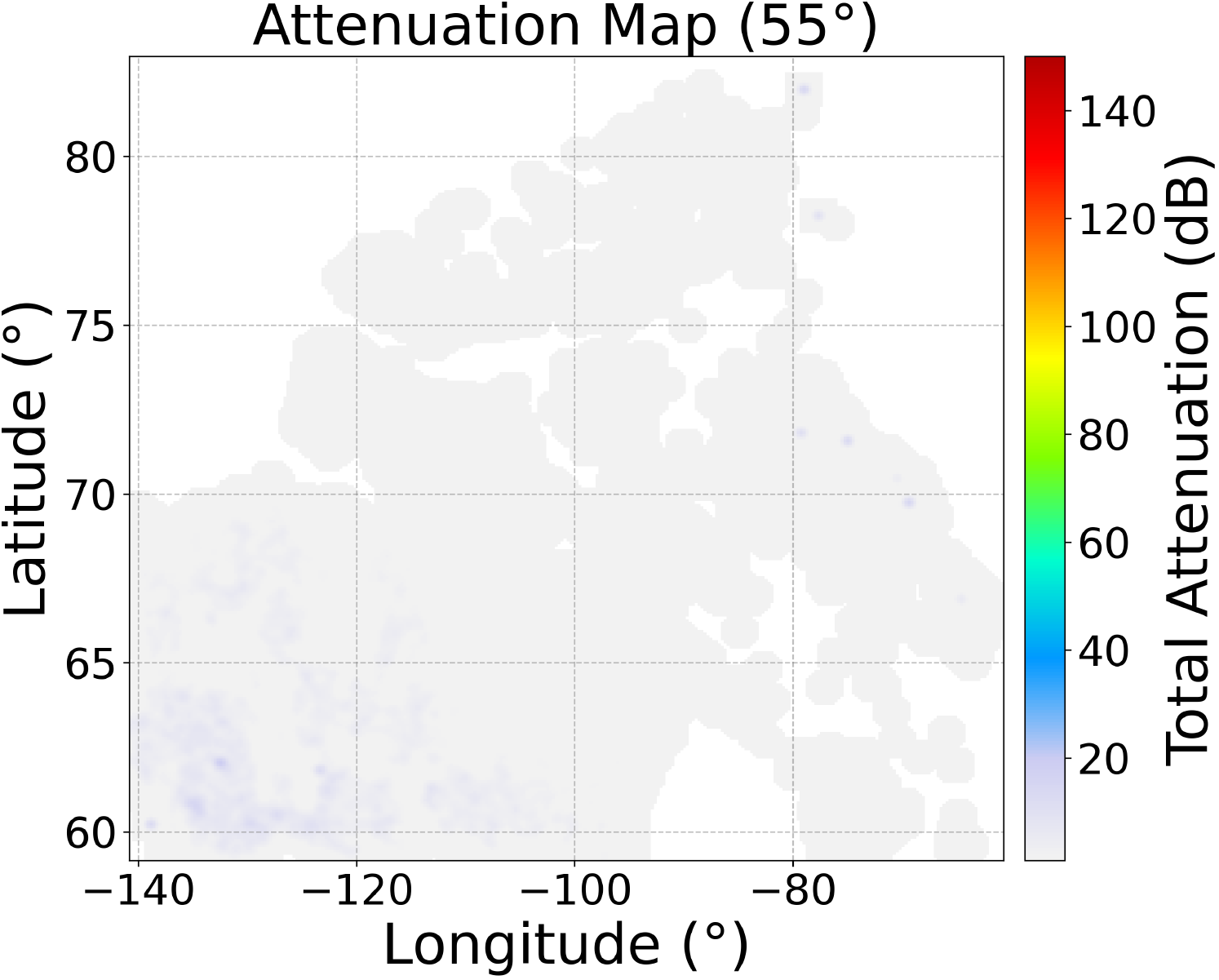}%
        \label{loss3}%
    }\\[0.5ex]

    \subfloat[$70^\circ$]{%
        \includegraphics[width=0.15\textwidth]{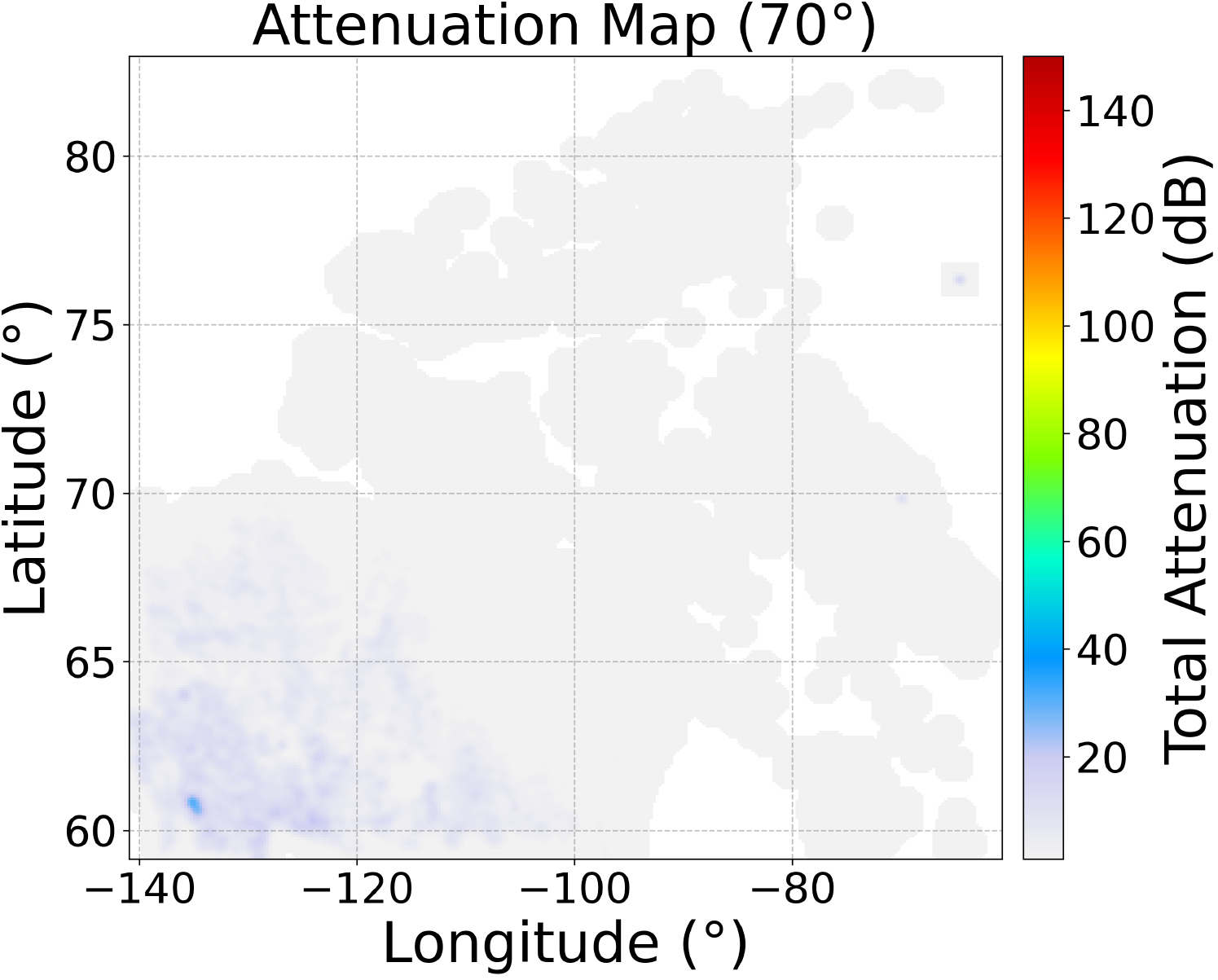}%
        \label{loss4}%
    }\hfil
    \subfloat[$85^\circ$]{%
        \includegraphics[width=0.15\textwidth]{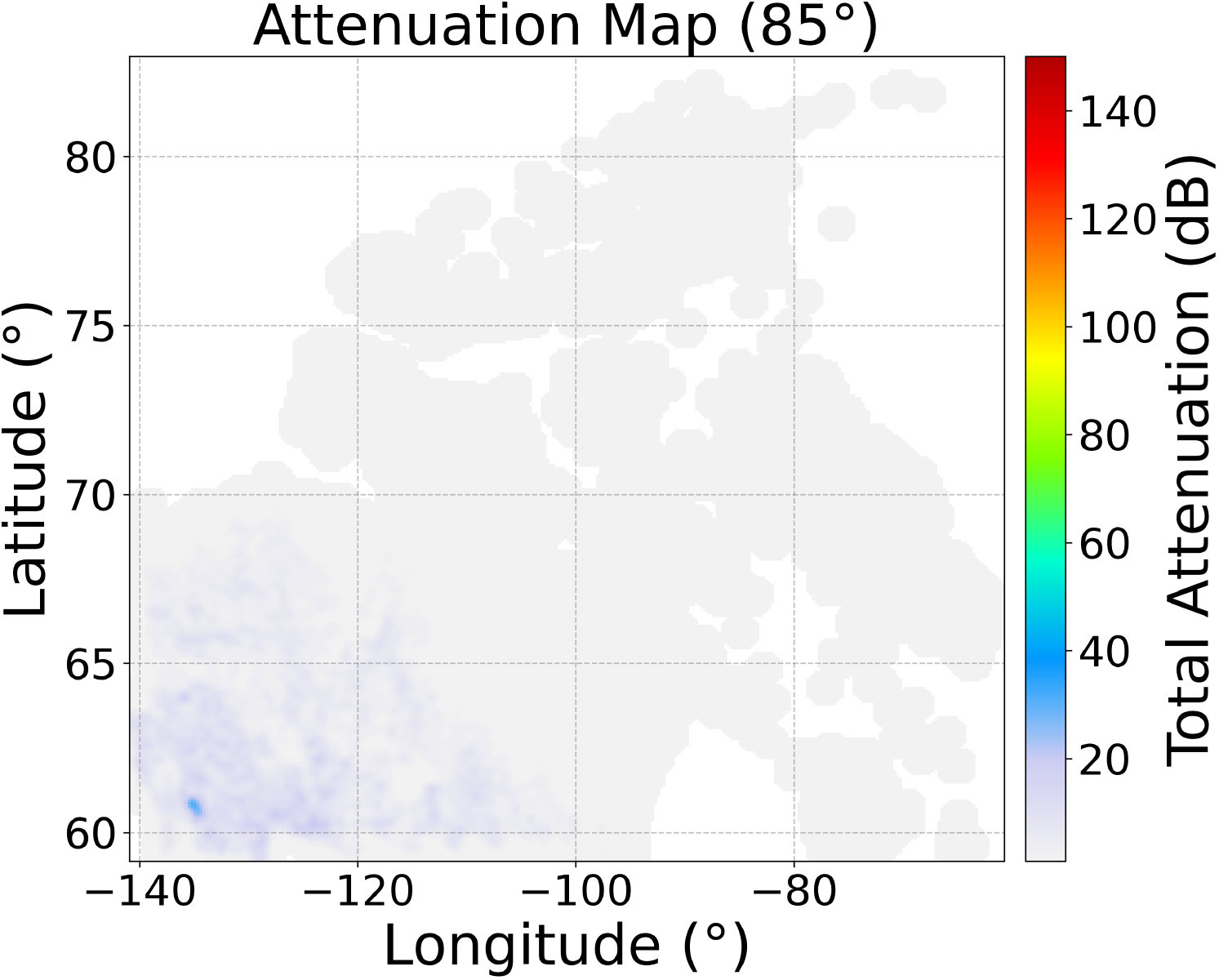}%
        \label{loss5}%
    }

    \caption{Attenuation heat map at different elevation.}
    \label{loss}
\end{figure}

Fig. \ref{loss} further shows the distribution of channel attenuation at different elevation angles. The results indicate that attenuation is concentrated in the northern and southern parts of the ROI, while it is negligible in the central region. This spatial heterogeneity reveals the regionalized impact of environmental factors on channel quality and provides a quantitative basis for resource scheduling and link management. Consequently, systems can anticipate coverage of high-attenuation regions and proactively implement dynamic resource allocation and optimization, thereby enhancing overall network performance and service reliability.

\subsection{Validation}
To validate the effectiveness of our method, we conducted evaluations using real-world data from Starlink \cite{starlink}, OneWeb \cite{oneweb}, and cellular networks. We discuss the results for each validation set in the following sections.

\subsubsection{Starlink}
As Starlink is currently the most mature satellite constellation, we compared model-estimated signal attenuation with real measured Starlink UT ping latency from the LENS dataset \cite{lens}, smoothed with a 60-second moving average to better observe overall trends and mitigate random noise. We use the Seattle subset of the LENS dataset, which shows noticeable obstruction despite UTs usually being deployed in open sites, making it representative. DEM \cite{us-dem} and land cover data \cite{us-lc} of Seattle were then used to estimate the channel condition at the UT location.

\begin{figure}
    \centering
    \includegraphics[width=0.9\linewidth]{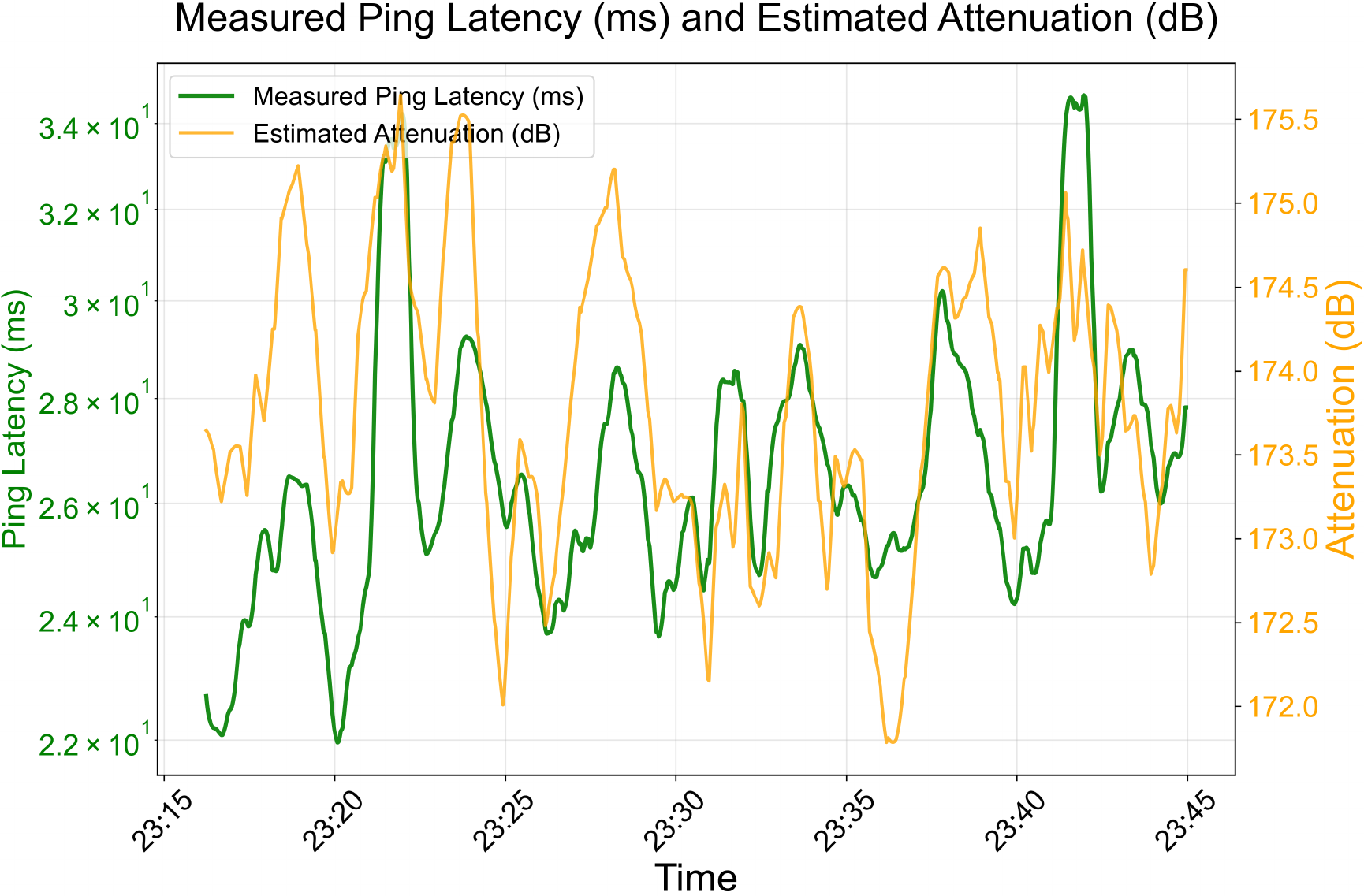}
    \caption{Validation on Starlink.}
    \label{satrlink}
\end{figure}

Since Starlink dishes do not provide access to channel quality measurements, we must rely on end-to-end Ping measurements to infer channel performance. In Fig. \ref{satrlink}, the yellow curve shows model-estimated signal attenuation (dB) and the green curve shows measured ping latency (ms) over 30 minutes. The changing signal attenuation was mainly due to the mobility of the satellite, resulting in changing elevation angles and obstructions. Though not directly comparable, Ping message delay reflects channel behavior and both exhibit consistent trends. To quantify this, we employed \textit{Sign Agreement}, since the two metrics differ in units and scales. It measures the proportion of instances where estimated ($x_i$) and measured ($y_i$) values change in the same direction. The correlation coefficient of $\boxed{\textbf{0.734}}$ indicates a strong correlation.
\begin{equation}
\text{Sign Agreement} = \frac{1}{N} \sum_{i=1}^{N} \mathbf{1}\big( \text{sign}(x_i) = \text{sign}(y_i) \big),
\end{equation}

Direct verification with Starlink is challenging, since its per-satellite downlink is characterized in terms of aggregate capacity, while phased-array beamforming dynamically reallocates beams and capacity to match demand \cite{starlink-fcc}, so the observed ping latency is affected by scheduling and beam management in addition to the physical channel state, and thus cannot reliably represent instantaneous channel conditions.

\subsubsection{OneWeb}
\begin{figure}
    \centering
    \includegraphics[width=0.9\linewidth]{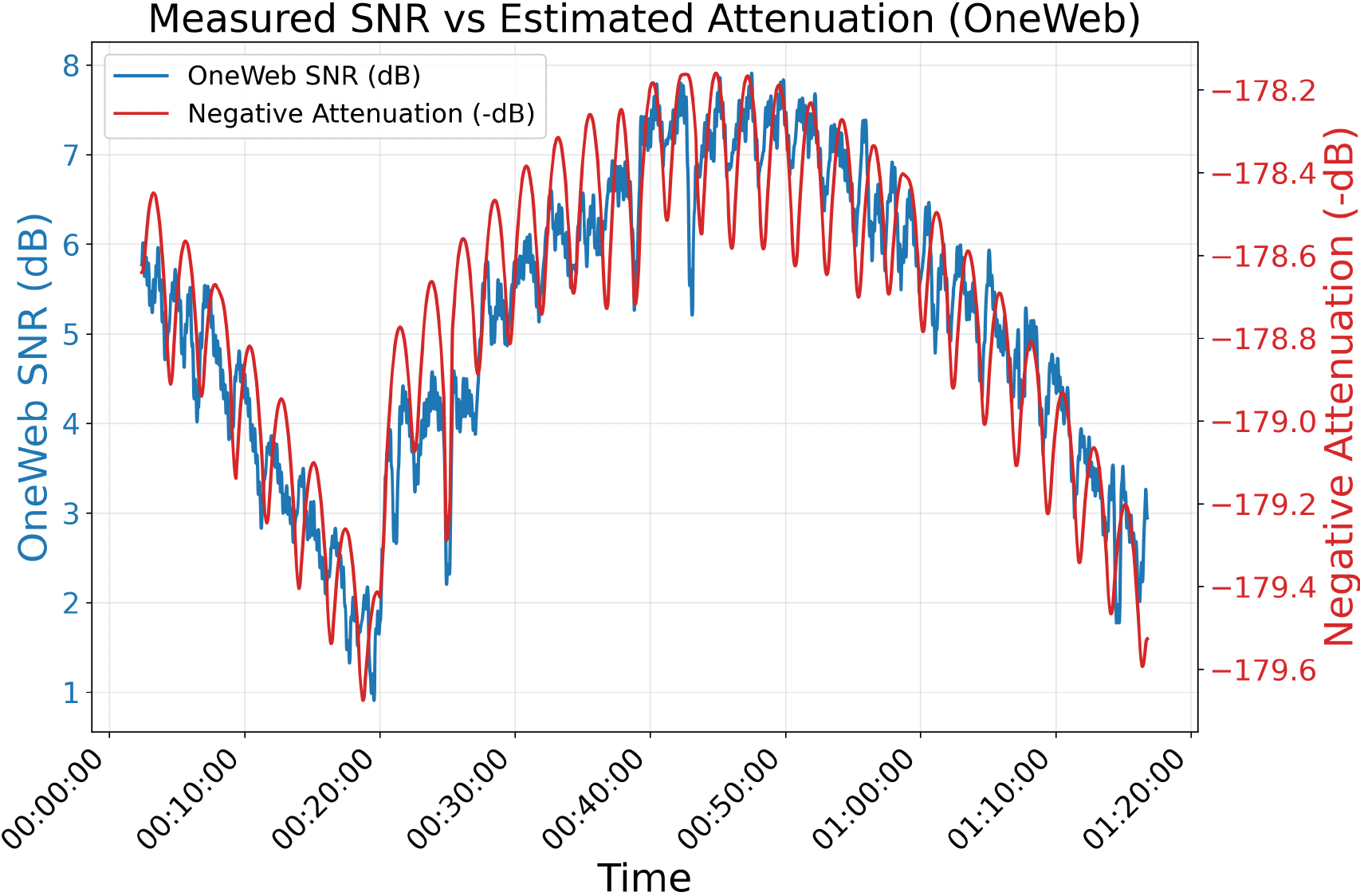}
    \caption{Validation on OneWeb.}
    \label{oneweb}
\end{figure}

OneWeb serves as another key constellation for validation as it exposes SNR readings to UTs. To directly validate our channel model using this metric, we compared the model-estimated signal attenuation with the OneWeb SNR collected in \cite{oneweb-val}. We applied our model to the environmental data \cite{us-dem, us-lc} at corresponding locations, and the results are illustrated in Fig. \ref{oneweb}. A strong correlation is observed between the OneWeb SNR (blue curve, dB) and the estimated attenuation (red curve, dB). Noting that attenuation and SNR are negatively correlated—where higher attenuation corresponds to lower SNR—we inverted the estimated attenuation values to align the visual trends. Given that both metrics are in the same dimension (dB), we employed the \textit{Pearson} correlation coefficient to quantify their linear correlation.

\begin{equation}
r = \frac{\sum_{i=1}^{N} (x_i - \bar{x})(y_i - \bar{y})}
{\sqrt{\sum_{i=1}^{N} (x_i - \bar{x})^2} \sqrt{\sum_{i=1}^{N} (y_i - \bar{y})^2}},
\end{equation}
where $x_i$ and $y_i$ denote the predicted and measured values, $\bar{x}$ and $\bar{y}$ their means, and $N$ the number of samples. Statistical significance is assessed by the p-value from the $t$-distribution with $N-2$ degrees of freedom, where $p<0.05$ is generally considered significant. Our correlation coefficient of $\boxed{\textbf{0.906}}$ with $p=0$ shows strong consistency between estimated and measured values.

\subsubsection{Cellular}
\begin{figure}
    \centering
    \includegraphics[width=0.9\linewidth]{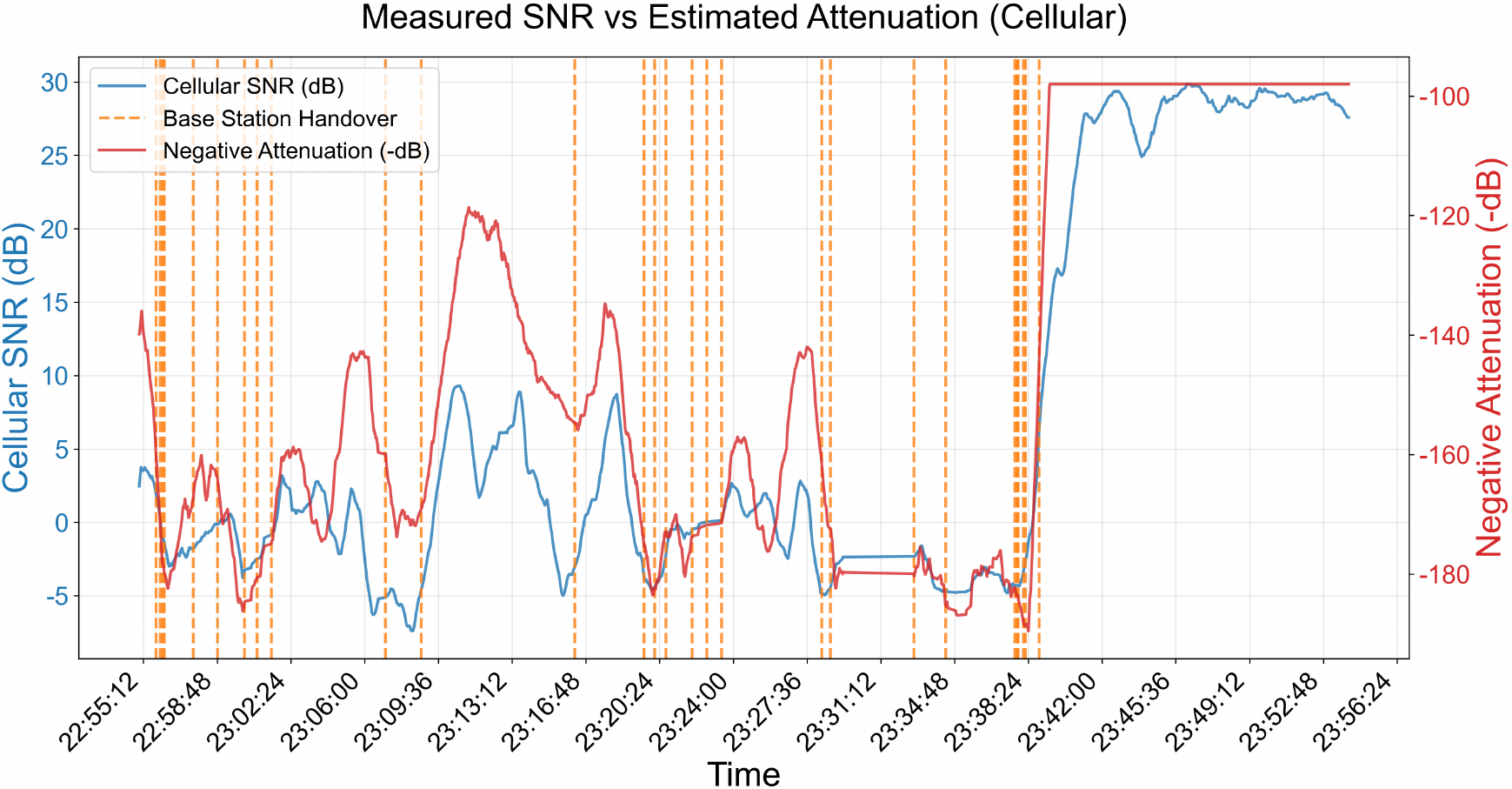}
    \caption{Validation on Cellular.}
    \label{cellular}
\end{figure}

\begin{figure}
    \centering
    \includegraphics[width=0.8\linewidth]{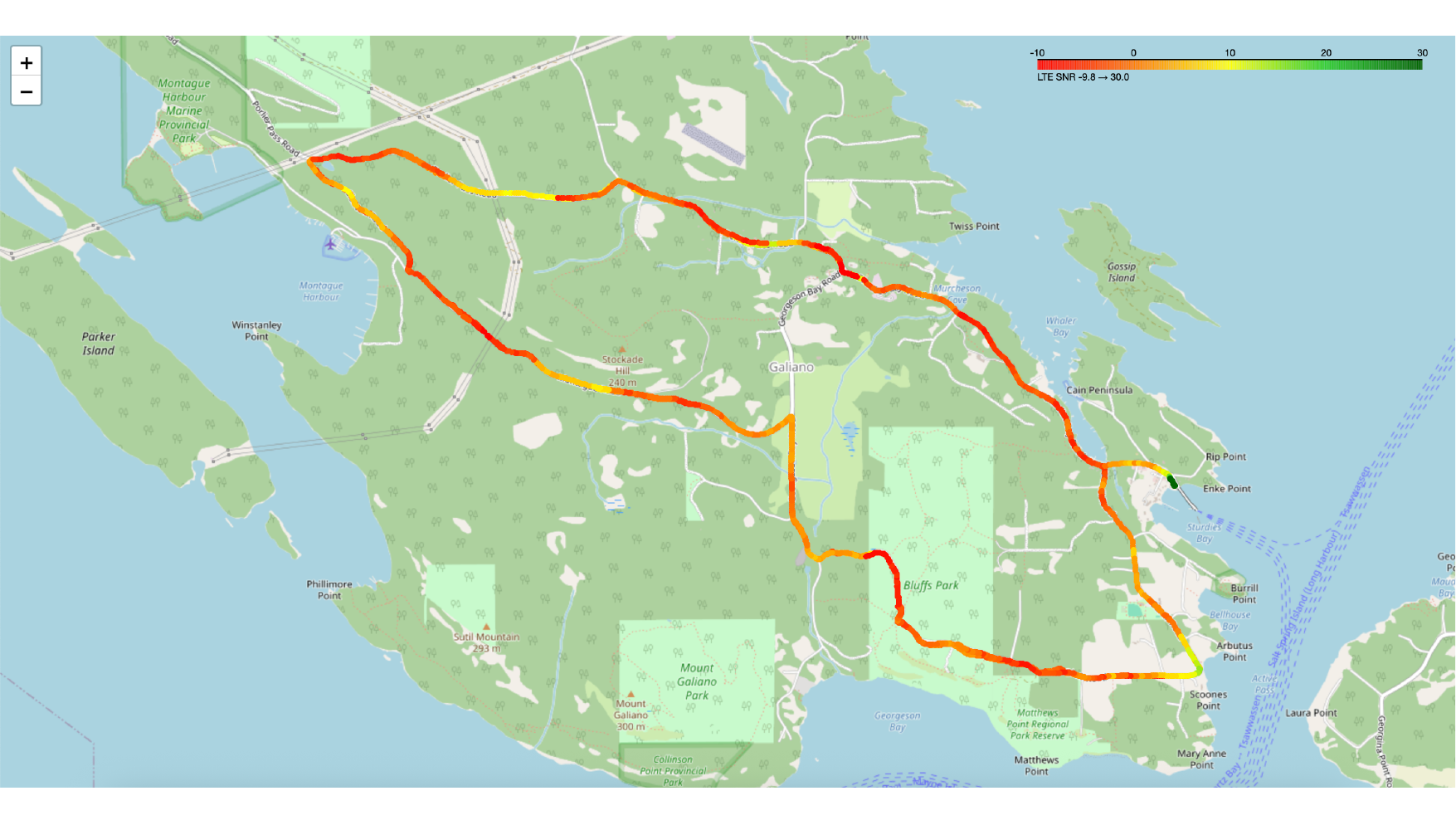}
    \caption{Trajectory Map.}
    \label{trajectory_map}
\end{figure}

We conducted cellular tests on Galiano Island, Canada, collecting LTE SNR while traveling \cite{finn}. Using DEM \cite{vic-dem} and land cover data \cite{lc}, our method estimated channel conditions at the corresponding GPS locations. Fig. \ref{cellular} shows the results: the red curve is estimated loss (dB), the blue curve is measured LTE SNR (dB), and the vertical orange lines mark base station handovers. Fig. \ref{trajectory_map} shows the actual driving route along which the measurements were collected. The \textit{Pearson} correlation coefficient of $\boxed{\textbf{0.932}}$ with $p=0$ shows strong consistency between estimated and measured values. While the measurements are affected by user behavior and base station power differences, the results sufficiently validate the feasibility of the proposed method.

\subsection{Prediction Accuracy of Diffusion Model}
To assess the capability of the proposed diffusion model in reconstructing full-field channel maps within complex geographical environments, this section evaluates prediction performance across varying degrees of observation sparsity. Given the highly complex and significantly heterogeneous surface and geometric features of the large-scale ROI, direct training on the entire area risks introducing data redundancy and compromising result interpretability. Consequently, we selected three representative and distinct 30 × 30 km sub-regions with distinct features: 

\begin{itemize}
    \item Region A: flat terrain with a standard deviation of \textbf{18.2 m} in elevation and essentially no vegetation, representative of open, unobstructed land.

    \item Region B: moderate relief with an elevation standard deviation of \textbf{563.2 m} and approximately 29.9\% vegetation coverage, capturing hilly areas with a mix of obstructed and unobstructed paths due to both terrain undulation and forested regions.

    \item Region C: rugged terrain with an elevation standard deviation of \textbf{701.8 m} and no vegetation, representing steep mountain regions.
\end{itemize}

To evaluate different levels of observation sparsity, we sample approximately 1\% and 4\% of the pixels in each sub-region. Our model was then applied to these sampled points to obtain the raw observations. For convolutional stability, subregions are formed into 256$\times$256 pixel tiles, where channel-loss observations are georeferenced with environmental features and satellite geometry. The model is trained for 500 epochs (batch size = 16) with a 250-step diffusion process.

\begin{figure}[htbp]
    \centering
    \subfloat[sample rate = 1\%.]{%
        \includegraphics[width=0.23\textwidth]{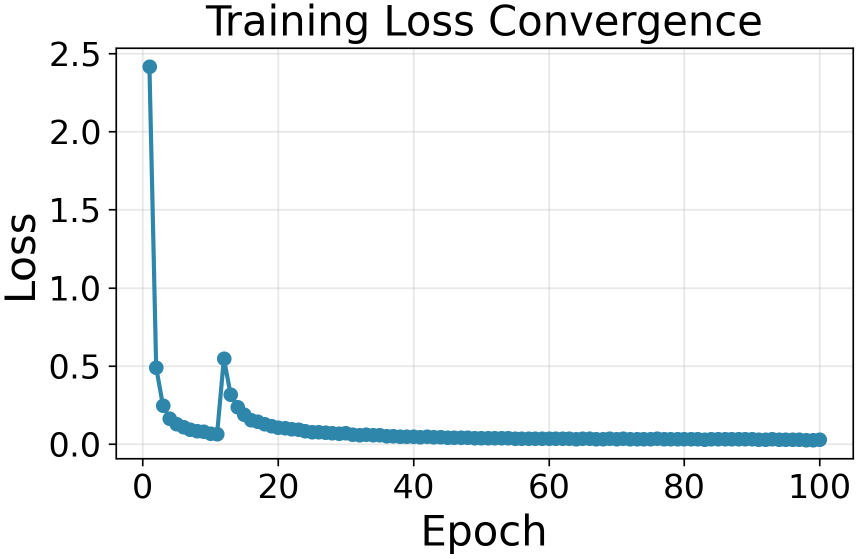}%
        \label{loss_convergence_1}%
    }\hfil
    \subfloat[sample rate = 4\%.]{%
        \includegraphics[width=0.23\textwidth]{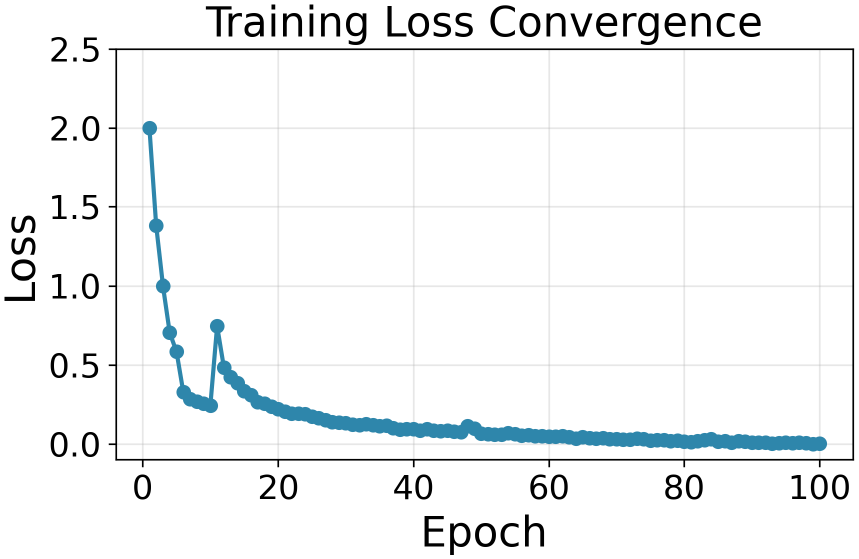}%
        \label{loss_convergence_4}%
    }
    \caption{Convergence curve.}
    \label{loss_convergence}
\end{figure}

Fig. \ref{loss_convergence} illustrates the loss convergence curve of the model over the first 100 training epochs. As observed, there is a rapid decline in training loss during the initial phase. This is attributed to the implementation of a \textit{Curriculum Learning} strategy: for the first 10 epochs, the model is fine-tuned exclusively on low-noise intervals, after which the training is extended to the full range of 250 noise diffusion steps starting from the 11th epoch. This design aims to prevent the model from being exposed directly to high-variance pure noise images during the parameter initialization stage, which could cause it to degenerate into a trivial solution of merely outputting the statistical mean to minimize global error.

Consequently, a transient rise in loss occurs at the 11th epoch upon the transition in difficulty level. Nevertheless, this is a normal adjustment following the increase in task complexity and does not affect the overall convergence trend, demonstrating the model's robust optimization stability. Notably, while the curve appears to descend more rapidly when using only sparse data (Fig. \ref{loss_convergence_1}), this is essentially an artifact indicative of overfitting, suggesting the model lacks the genuine generative capacity to handle complex variations. Conversely, the fluctuations observed in the convergence curve with denser observations (Fig. \ref{loss_convergence_4}) reflect the model's transition from "rote memorization" to a true "understanding" of the feature distribution.

\begin{figure}
    \centering
    \includegraphics[width=0.8\linewidth]{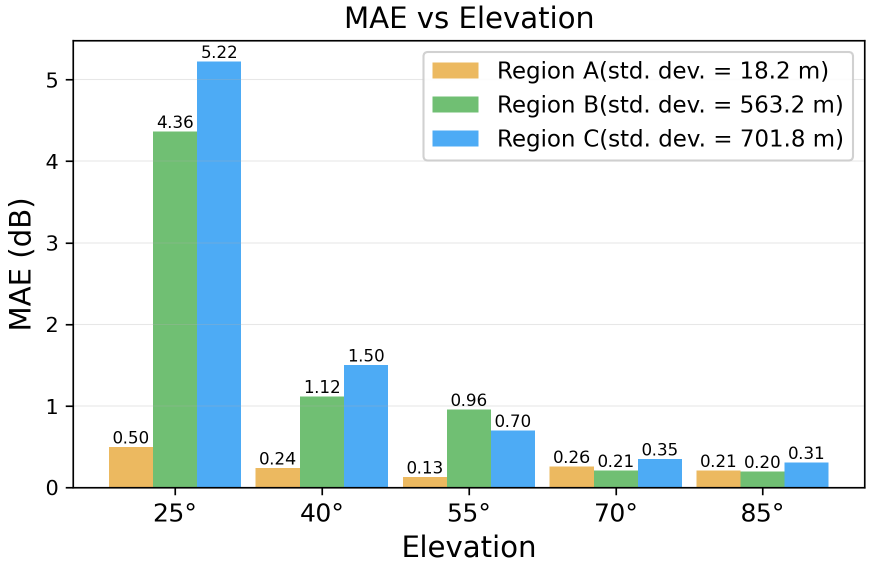}
    \caption{Diffusion model performance when sample rate = 1\%.}
    \label{mae_10}
\end{figure}

\begin{figure}
    \centering
    \includegraphics[width=0.8\linewidth]{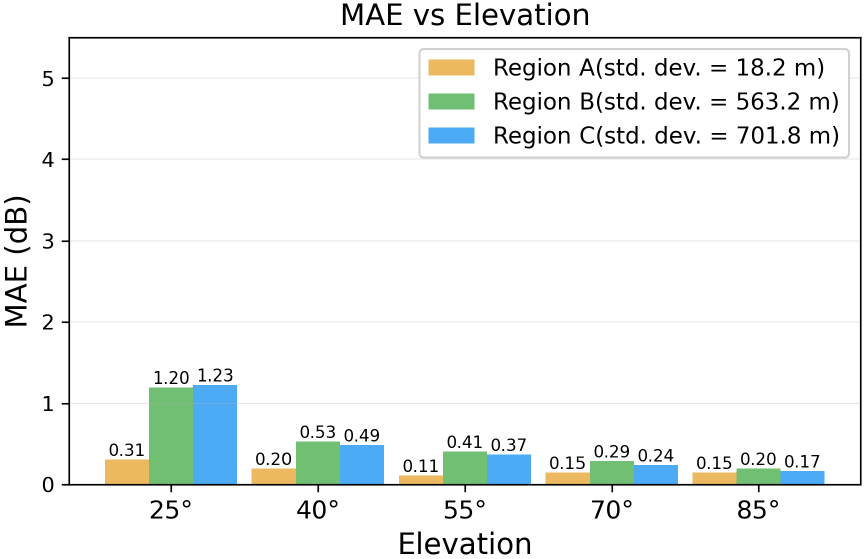}
    \caption{Diffusion model performance when sample rate = 4\%.}
    \label{mae_5}
\end{figure}

Fig. \ref{mae_10} and Fig. \ref{mae_5} compare the predicted Mean Absolute Error (MAE) across three sub-regions under different sampling densities: a sampling rate of approx. 1\% and a sampling rate approx. 4\%, respectively. It shows that increasing the sampling density significantly reduces prediction errors, particularly in scenarios characterized by low elevation angles and rugged terrain. For instance, in Region C, the MAE drops drastically from 5.22 dB to 1.23 dB — a reduction of approximately 76.4\%. In contrast, at high elevation angles, the link is nearly vertical and thus less susceptible to environmental obstructions, the MAE in these cases remains consistently low and is less sensitive to changes in sampling density. 

We further compare the computational efficiency between the RT method and the diffusion-based inference. For a 256 × 256-pixel tile (65,536 pixels in total), RT computation requires roughly $\boxed{\textbf{144000 seconds}}$, whereas the diffusion model completes forward inference in about $\boxed{\textbf{300 seconds}}$. Overall, despite small accuracy degradation, it substantially reduces the total computational cost. Results in Fig. \ref{mae_10} and Fig. \ref{mae_5} also indicate that the extent of pre-calculation can be flexibly determined based on the specific accuracy requirements of different tasks.

\section{Conclusion}
This paper presents an environment-aware air-to-ground channel modeling method that integrates terrain, land cover, and climate data to estimate channels, validated with real-world measurements. A diffusion model further predicts channel states for arbitrary UT–satellite pairs, reducing reliance on ray tracing and making our solution more scalable. The proposed model can support resource scheduling and link management, offering practical value for network deployment and optimization. For future work, we will refine the model to better capture seasonal vegetation dynamics, validate it across diverse environments, and incorporate it into LEO network resource management and enhance handover performance. 

\section*{Acknowledgment}
The authors would like to thank Dr. Jianping Pan, Jinwei Zhao and Finnegan Archer for their assistance in the collection of real-world measurement data.

\newpage





\vfill

\end{document}